\documentclass[twocolumn,aps,amsmath,amsfonts,amssymb]{revtex4-1}

\usepackage{graphicx}
\usepackage{amsmath,amsfonts,amssymb,color,bbm}
\usepackage{enumerate}
\usepackage{subfigure}

\newcommand{\abs}[1]{\left\lvert #1 \right\rvert}
\renewcommand{\vec}[1]{\mathbf{#1}}
\renewcommand{\d}{\mathrm{d}}
\newcommand{\ez}{{1,2}}
\newcommand{\ze}{{2,1}}

\usepackage{xfrac}

\begin{document}

\title{Coupled superfluidity of binary Bose mixtures in two dimensions}
\author{Volker Karle}
\email[Currently at IST\,Austria. Email:\,]{vkarle@ist.ac.at}
\affiliation{Institute for Theoretical Physics, Heidelberg University, D-69120 Heidelberg, Germany}

\author{Nicol\`o Defenu}
\affiliation{Institute for Theoretical Physics, Heidelberg University, D-69120 Heidelberg, Germany}

\author{Tilman Enss}
\affiliation{Institute for Theoretical Physics, Heidelberg University, D-69120 Heidelberg, Germany}

\begin{abstract}
We consider a two-component Bose gas in two dimensions at low temperature with short-range repulsive interaction. 
In the coexistence phase where both components are superfluid, inter-species interactions induce a nondissipative drag between the two superfluid flows (Andreev-Bashkin effect). We show that this behavior leads to a modification of the usual Berezinskii-Kosterlitz-Thouless (BKT) transition in two dimensions. We extend the renormalization of the superfluid densities at finite temperature using the renormalization group approach and find that the vortices of one component have a large influence on the superfluid properties of the other, mediated by the nondissipative drag. The extended BKT flow equations indicate that the occurrence of the vortex unbinding transition in one of the components can induce the breakdown of superfluidity also in the other, leading to a locking phenomenon for the critical temperatures of the two gases. 
\end{abstract}
\date{\today}
\maketitle

\section{Introduction}
The physics of degenerate two-component bosonic mixtures plays an important role in various systems, such as $^3$He-$^4$He mixtures~\cite{larsen1963binary, graf1967phase, andreev1975af}, 
ultracold atomic gases of different species or hyperfine states \cite{ho1996binary, myatt1997production,  altman2003phase, stamper2013spinor}, bilayer Bose systems~\cite{fil2004drag} or two-gap superconductors~\cite{szabo2001evidence}. Recently, experimental progress in creating degenerate two-component Bose droplets and mixtures~\cite{petrov2015quantum, cheiney2018bright, cabrera2018quantum, ye2018dressed, schulze2018feshbach} has raised the question whether superfluidity is robust and how the superfluid behavior in each component is influenced by the presence of the other. Furthermore, while the behavior of low dimensional Bose mixtures in the quantum degenerate regime has already been established~\cite{petrov2000regimes, petrov2016ultradilute}, the finite temperature picture for two coupled components has yet to be clarified.

In the quantum degenerate case, the inter-species interaction leads to the hybridization of the low-energy phonon excitations of the two components~\cite{busch1997stability, bashkin1997instability, ao2000two, pethick2002bose, pitaevskii2016bose}. The superfluid behavior is mediated by the long-wavelength phonon fluctuations of the mixture, and as a result a nondissipative drag or Andreev-Bashkin interaction between the superfluid flows of both components appears~\cite{andreev1975af, fil2005nondissipative, ishino2011countersuperflow, hofer2012superfluid, nespolo2017andreev, svistunov2015superfluid, parisi2018spin, sellin2018, Konietin2018}. 
For the single-component Bose gas in two dimensions at finite temperature, superfluidity is eventually destroyed by topological vortex excitations which drive the BKT transition to the normal state \cite{berezinskii1972destruction, kosterlitz1973ordering, kosterlitz1974critical}. What is the role of topological excitations in the Bose mixture \cite{dahl2008}, and what effect do they have on the nondissipative drag? Our results suggest that the interaction between different topological excitations can lead to coupled superfluidity in both components, with both critical temperatures locked to a unique value.

In this paper we compute the phase diagram of the degenerate two-component Bose mixture with equal mass first on the basis of  long-wavelength phonon fluctuations, and quantify the nondissipative drag originating from the these fluctuations at zero and finite temperature. We then build on these results to estimate the effect of vortex excitations on the superfluid behavior. Specifically, we extend the BKT renormalization group (RG) flow equations from one to two components, and explicitly include the interaction effect between topological excitations of different components. We find a strong renormalization of the superfluid densities and the nondissipative drag due the topological excitations. Remarkably, this can even lead to the breakdown of superfluidity in \emph{both} components as soon as topological excitations become large in one component. This behavior could be observed in experiments which image vortices \cite{hadzibabic2006berezinskii}.

Our findings have important consequences for other two-components systems which can be mapped to Bose mixtures including species with different masses, such as Na and K. Besides ultracold atom experiments, this could be important for bilayer Bose systems~\cite{fil2004drag, nespolo2017andreev, rancon2017kosterlitz} and classical nonequilibrium simulations~\cite{takeuchi2010binary, karl2013universal, karl2017strongly}, but also in connection with the spatial form of vortices within a Bose mixture~\cite{nespolo2017andreev, gallemi2018magnetic} and for
two-component Bose mixtures with inter-component Josephson coupling~\cite{kobayashi2018berezinskii}. Furthermore, in two-band superconductor such as MgB$_2$ \cite{szabo2001evidence} with multiple energy gaps, different types of Cooper pairs can form a binary Bose mixture~\cite{svistunov2015superfluid}. It was also proposed that mixtures of neutron and proton Cooper pairs form a condensate inside  neutron stars~\cite{babaev2004andreev} and within the metallic state of hydrogen~\cite{babaev2005observability}.

Furthermore, several fundamental condensed matter problems are related to the physics of bilayer models and coupled field theories\,\cite{LeClair1998}, such as heavy fermion systems\,\cite{Stewart1984} and the prototypical Kondo lattice model\,\cite{Coleman2015}, whose critical properties have been connected to the physics of coupled quantum spin chains\,\cite{Strong1994}. More recently, much attention has been devoted to twisted bilayer graphene\,\cite{Po2018,Ramires2018,Peltonen2018} where experimental evidences of superconductivity have been observed\,\cite{Cao2018}.

The present analysis can also be cast in the wider framework of layered two dimensional systems, which are deeply connected to high-temperature superconductors\,\cite{Leggett2006}. In this context, generalized BKT flow equations have been derived\,\cite{Pierson1994,Nandori2005,mathey2008} and successfully applied to the description of transport properties in strongly correlated superconductors\,\cite{Pierson1995,Nandori2007}.

The paper is structured as follows. In Sec.~\ref{sec:mf} we introduce the model and show its mean-field phase diagram, in Sec.~\ref{sec:ex} we present the mixed phonon modes and the phase diagram of the superfluid densities without topological excitations. In Sec.~\ref{sec:top_exc} we derive the new coupled RG flow which shows how the topological excitations alter the superfluid behavior in 2D. We discuss the implications in Sec.~\ref{sec:results} and conclude with Sec.~\ref{sec:conclusion}.

\section{Mean-field phase diagram}
\label{sec:mf}
\begin{figure}[t]
    \centering
    \includegraphics[width=1.0\linewidth]{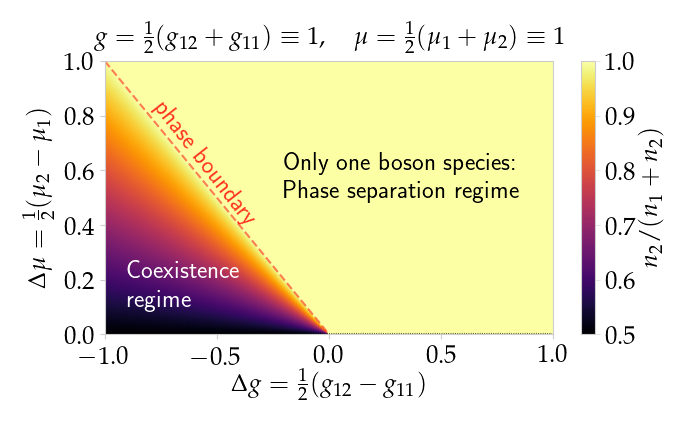}
    \caption{Mean-field phase diagram for a Bose mixture in the ground state.  A mixture occurs only in the coexistence regime on the left side where the inter-species repulsion is smaller than intra-species repulsion (on the left margin $\Delta g = -g$, $g_{12}=0$ and the inter-species repulsion vanishes).  In contrast, in the phase separation regime there are only bosons of species 2 because $\mu_2>\mu_1$.  In this work, we study how fluctuations modify the coexistence regime.  Note that we have chosen units such that $g=1$ and $\mu=1$.}
\label{fig:phasediagram}
\end{figure}

We consider a weakly interacting binary Bose mixture with equal masses in two dimensions, which is described by the Lagrangian \cite{ho1996binary, pethick2002bose, pitaevskii2016bose}
\begin{multline}
    \mathcal{L}(x,t) = \sum_{i}\psi_i^\dagger(x,t) \Bigl[i \partial_t -\frac{\nabla^2}{2m} - \mu_i \Bigr] \psi_i(x,t) \\ 
    + \frac{1}{2}  \sum_{ij}g_{ij}\abs{\psi_i(x,t)}^2\abs{\psi_j(x,t)}^2.
    \label{eq:L1}
\end{multline}
Here, $\psi_i(x,t)$ denote complex bosonic fields for species $i=1,2$ with equal masses $m$ but individual chemical potentials $\mu_i$. The short-range interaction is assumed repulsive both within species ($g_{11},g_{22}>0$) and between species ($g_{12}>0$).  The coupling strengths $g_{ij}$ are given in terms of the physical 2D scattering lengths $a_{ij}>0$~\cite{adhikari1986quantum, petrov2000bose, petrov2001interatomic, salasnich2016zero, Konietin2018},
\begin{equation}
    g_{ij}(E) = \frac{4\pi/m}{\ln(4/e^{2\gamma} m a_{ij}^2E)}\,.
    \label{eq:gE12}
\end{equation}
Note that the coupling strength always depends on the scattering energy $E$, for which we insert the chemical potential $\mu$ as the typical many-body energy scale to incorporate the effect of quantum fluctuations (see Appendix~\ref{sec:phonon} for a discussion).  From now on we restrict ourselves to the case of equal intra-species scattering $g_{11} = g_{22}$.  An asymmetry between $\mu_1,\mu_2$ or $g_{11},g_{12}$ then leads to nonsymmetric superfluid behavior.  In terms of the number densities $n_i(x,t) = \abs{\psi_i(x,t)}^2$ we arrive at the Lagrangian for the potential part
\begin{equation}
    \mathcal L_\text{pot} = -\mu_1 n_1 - \mu_2 n_2 + \tfrac12 g_{11} (n_1^2 + n_2^2) + g_{12}n_1 n_2.
\label{eq:L2}
\end{equation}
We introduce relative variables $\Delta \mu = \tfrac{1}{2}(\mu_2 - \mu_1), \, \mu = \frac{1}{2}(\mu_2 + \mu_1),\, \Delta g = \tfrac{1}{2}(g_{12} - g_{11}), \, g = \tfrac{1}{2}(g_{12} + g_{11})$ and $n = n_1 + n_2,\, \Delta n = n_2 - n_1$ to write
\begin{equation}
\begin{aligned}
    \mathcal{L_\text{pot}} &= \frac{1}{2} g n^2  - \frac{1}{2} \Delta g {(\Delta n)}^2  - \mu n - \Delta \mu \Delta n\\ & = \tfrac{g}{2}(n - \tfrac{\mu}{g})^2 - \tfrac{\Delta g}{2}(\Delta n + \tfrac{\Delta \mu}{\Delta g})^2 + \text{const},
\end{aligned}
\label{eq:Lpot}
\end{equation}
where the second line holds for $g,\Delta g \neq 0$. The potential is minimized by the mean-field solution $\bar n = \bar n_1+\bar n_2 = \mu/g$ and 
\begin{equation}
   \Delta \bar n = \bar n_2 - \bar n_1 = \begin{cases}
- \frac{\Delta \mu}{\Delta g} & \tfrac{\abs{\Delta \mu}}{\mu} \leqslant \tfrac{\abs{\Delta g}}{g} \text{ and } \Delta g < 0,\\
 \tfrac{\mu}{g}\, \text{sign}[\Delta \mu] & \text{otherwise.} 
\end{cases}
\label{eq:density_MF}
\end{equation}
The mean-field phase diagram exhibits a coexistence regime (miscible, first case) and a phase separation regime (immiscible, second case), as illustrated in Fig.~\ref{fig:phasediagram}.  In the following, we will focus on the left half of the diagram, i.e., for $\Delta g < 0$. For $\Delta \mu <0 $ the phase diagram is the mirror image with $\bar n_2 \leftrightarrow \bar n_1$. 

\section{Phonon excitations}
\label{sec:ex}
\begin{figure}[t]
    \centering
    \includegraphics[width=1.0\linewidth]{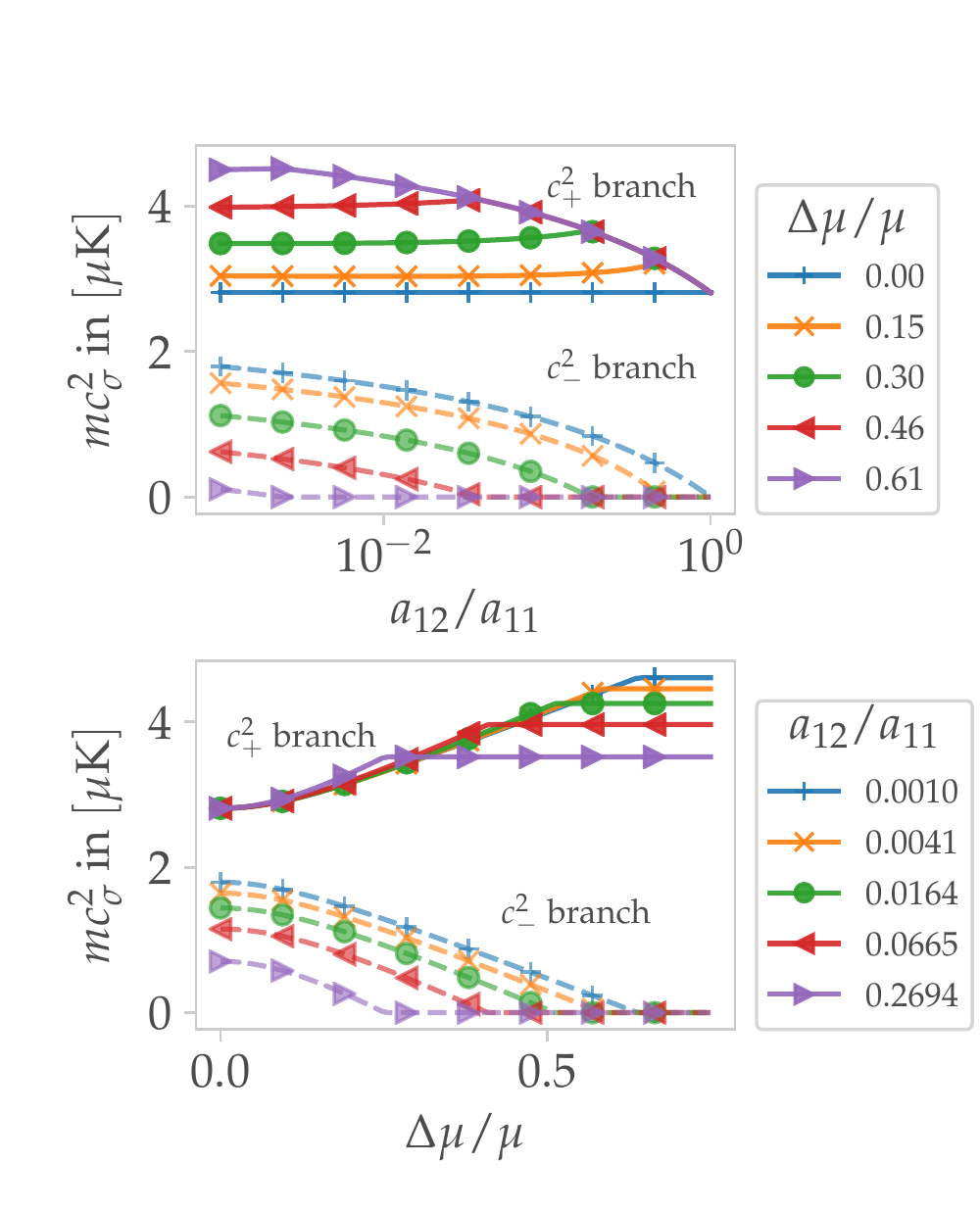}
    \caption{Speed of sound $c_\pm$ of two branches of normal modes from Eq.~\eqref{eq:cpm}. The lower branch $c_-\to0$ becomes soft at the transition to the phase separation regime where $n_1\to0$, while the upper branch (dashed line) $c_+\to c_2$. Parameters are for a $^{87}$Rb Bose mixture with scattering length $a_{11}=100 a_B$ and varying $a_{12}$. The energy scale $E$ in Eq.~\eqref{eq:gE12} is set to the many-body scale $\mu = (\mu_1+\mu_2)/2 = 2.8 \mu\mathrm{K}$, and we vary the chemical potential difference $\Delta \mu = (\mu_2 - \mu_1)/2$.}
\label{fig:sos}
\end{figure}
In two dimensions, a Bose-Einstein condensate with long-range coherence does not exist at any finite temperature. However, there can be a quasi-condensate with finite superfluid density $n_s$ \cite{svistunov2015superfluid,Prokofiev2004}. The low-energy fluctuations around the quasi-condensates in the mixture are given by two branches of phonon modes with dispersion \cite{ao2000two, fil2005nondissipative} (see Appendix~\ref{sec:phonon} for a derivation),
\begin{equation}
    \omega^2_{k\pm} = \epsilon_k (\epsilon_k + 2m c^2_\pm),
    \label{eq:phonon_freq}
\end{equation}
in terms of the speed of sound
\begin{multline}
   2 mc_\pm^2 = g_{11}(\bar{n_{1}}+\bar{n}_{2})\\
   \pm \sqrt{g_{11}^2  (\bar{n}_{2}-\bar{n}_{1})^2 + 4 g_{12}^2 \bar{n}_1 \bar{n}_2 },
   \label{eq:cpm}
\end{multline}
where we used the mean-field densities $\bar{n}_i$ defined in~\eqref{eq:density_MF}. The linearity of the dispersion relation allows for superfluid behavior~\cite{landau1980course}. The two branches of normal modes are combined excitations of components 1 and 2, corresponding to density ($c_+$) and spin ($c_-$) excitations \cite{abad2014persistent}, see Fig.~\ref{fig:sos}. For definiteness, we have chosen parameters for the experimentally relevant case of a $^{87}$Rb mixture in two hyperfine states. The spin mode becomes soft, $c_-\to0$, at the quantum phase transition to the phase separated regime, while the density mode $c_+\to c_2$ approaches the speed of sound of component 2.

The zero-point quantum fluctuations of the normal modes contribute to the ground-state equation of state as a shift of the energy density. In Appendix~\ref{sec:phonon} we show that this can be re-absorbed in a logarithmic correction to the coupling $g_{ij}$ in Eq.~\eqref{eq:gE12} \cite{petrov2016ultradilute, salasnich2016zero}.  

Unlike the single-component case, for two components we encounter a nondissipative drag, the Andreev-Bashkin entrainment effect~\cite{andreev1975af}, between the two superfluid currents $\vec{j}_s^{(i)}$ (see~\cite{svistunov2015superfluid} for a comprehensive introduction). The supercurrents $\vec{j}_s^{(i)} = (mL^2)^{-1}\d\Omega_\text{fl}/\d\vec{v}_i$ can be computed from the grand potential $\Omega_\text{fl}$ of long-wavelength phonon fluctuations, and are expressed in terms of the superflows $\vec{v}_i = m^{-1} \nabla \theta_i$ as
\begin{align}
    \vec{j}_1 &= (\bar{n}_1 - n_{n1} - n_\text{dr}) \vec{v}_1 + n_\text{dr} \vec{v}_2 = \tilde n_1 \vec{v}_1 + n_\text{dr}\vec{v}_2,\notag\\
    \vec{j}_2 &= (\bar{n}_2 - n_{n2} - n_\text{dr}) \vec{v}_2 + n_\text{dr} \vec{v}_1= \tilde n_2 \vec{v}_2 + n_\text{dr}\vec{v}_1.
    \label{eq:density_s}
\end{align}

\begin{figure}[t]
    \centering
    \includegraphics[width=1.05\linewidth]{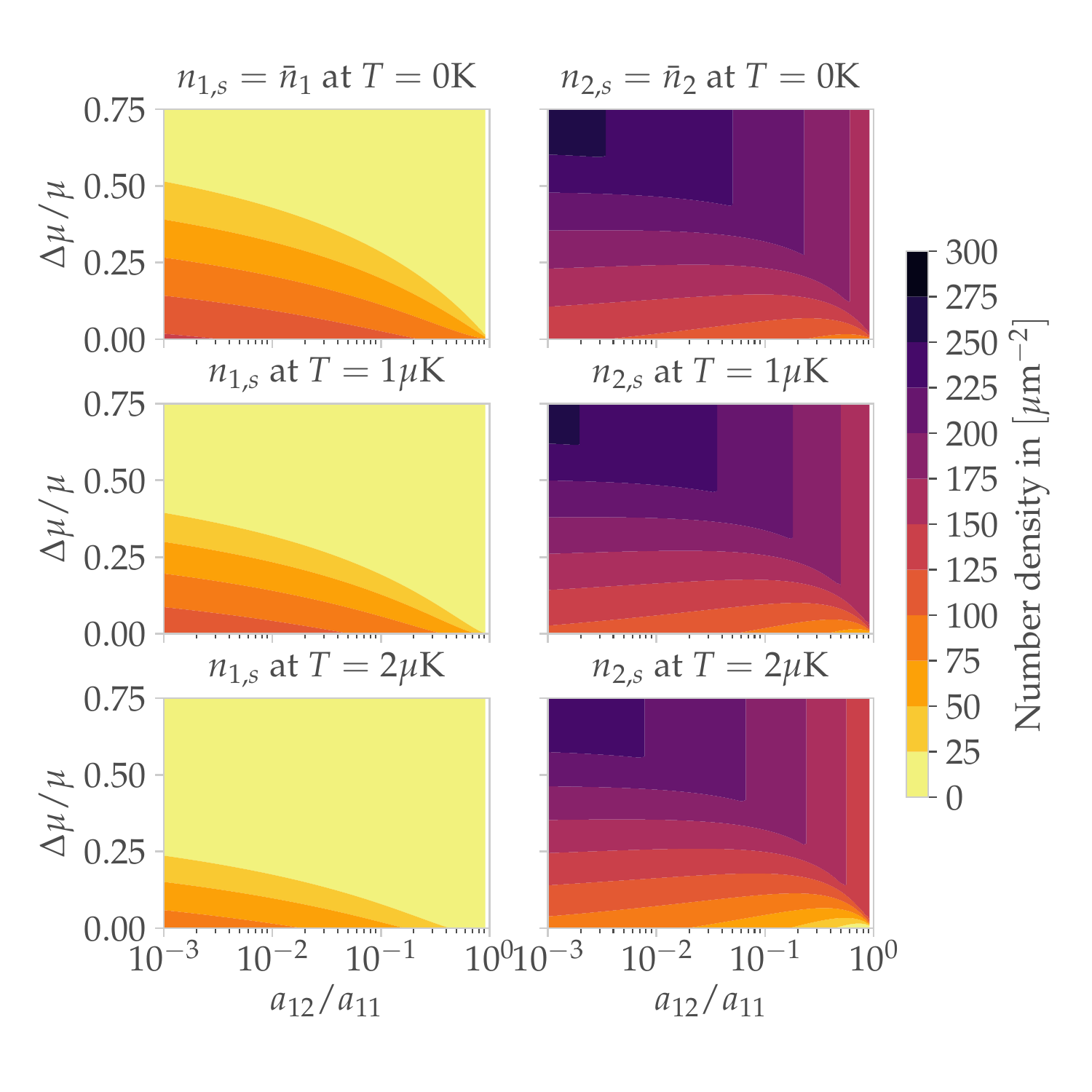}
    \caption{Depleted densities $n_{s,i} = \bar{n}_i - n_{n,i}$ at different temperatures from long-wavelength excitations only, at fixed chemical potentials $\mu_1,\mu_2$ (parameters for $^{87}$Rb as in Fig.~\ref{fig:sos}). At zero temperature both components are fully superfluid. The normal densities $n_{n,i}$ \eqref{eq:rho_ni} increase with temperature, and they become larger for softer phonons with smaller speed of sound \eqref{eq:cpm}. In the coexistence region, both phonon modes contribute to both normal densities; therefore, the fluctuations of one component decrease the superfluid density of the other component as well.}
\label{fig:superfluid_densities}
\end{figure}
\begin{figure}[t]
    \centering
    \includegraphics[width=1.05\linewidth]{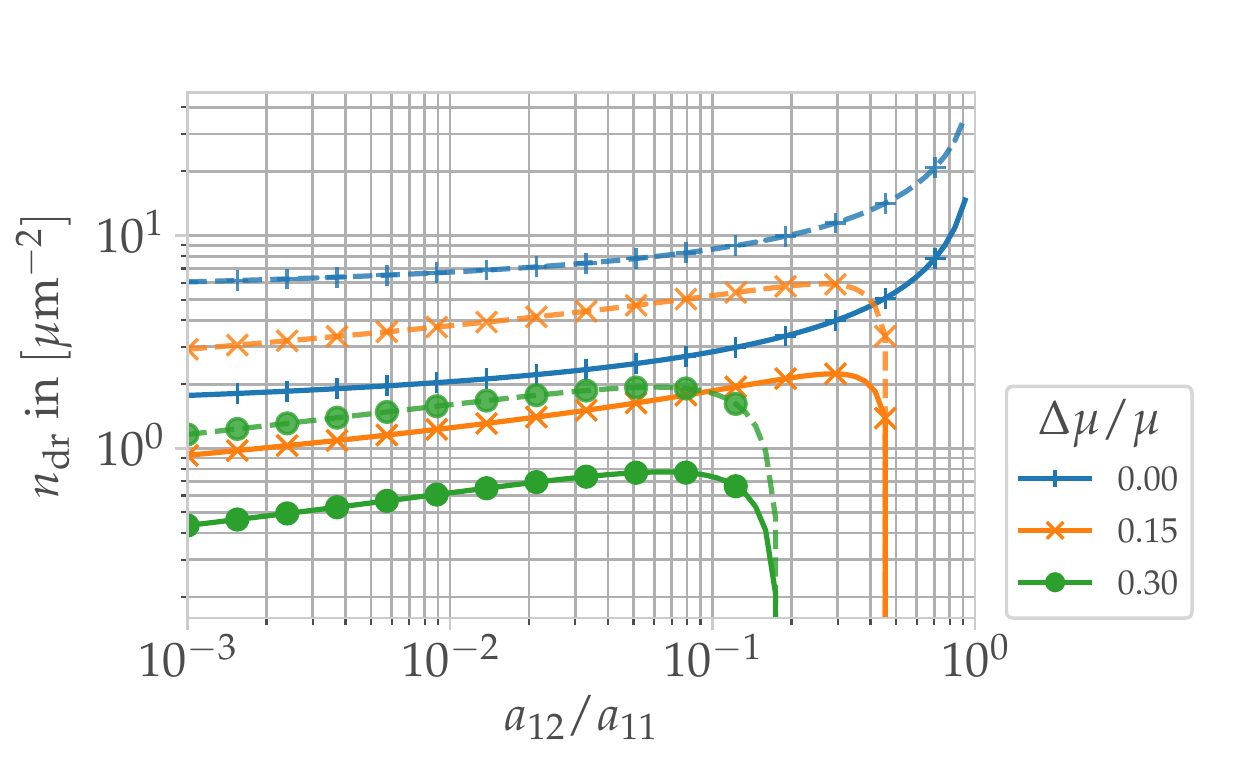}
    \caption{Nondissipative drag density $n_\text{dr}$ at different temperatures, $T=1\mu$K (solid line) and $T=2\mu$K (dashed line), from long-wavelength fluctuations; parameters for $^{87}$Rb as in Fig.~\ref{fig:sos}. Since the nondissipative drag arises from fluctuations between the two components, it becomes larger for increasing temperature and for decreasing speed of sound near the phase transition (cf.\ Fig.~\ref{fig:sos}). Finite values of the drag density $n_\text{dr} \propto \bar n_1  \bar n_2$ \eqref{eq:drag} are only possible in the coexistence phase, where both components are present, while the drag density trivially vanishes in the regime of phase separation.}
    \label{fig:drag_densities}
\end{figure}

Thermal and quantum fluctuations give rise to the temperature dependent drag density $n_\text{dr}$ as well as the normal densities $n_{n,i}$, which in turn define the depleted densities $n_{s,i} = \bar n_i - n_{n,i}$ (see Appendix~\ref{sec:andreev} for explicit expressions for $n_{n,i}$ and $n_\text{dr}$). The diagonal coefficients then give the superfluid densities $\tilde{n}_i = n_{s,i} - n_\text{dr}$. To demonstrate the quantitative importance of these fluctuation effects, we have computed both the superfluid densities (Fig.~\ref{fig:superfluid_densities}) and the drag density (Fig.~\ref{fig:drag_densities}). One observes that the fluctuations become large, and superfluidity is suppressed by the normal component, when approaching the quantum phase transition where spin modes become soft, $c_-\to0$.

From Eq.~\eqref{eq:density_s} we see that the classical action of the phase fluctuations can be described by the Villain model, which is found by assuming the simplest action quadratic in $\vec{v}_s$ which reproduces the given values of $n_s$~\cite{svistunov2015superfluid}. A more rigorous approach would be to use an RG treatment which includes density and phase fluctuations on equal footing~\cite{defenu2017nonperturbative}. The two-component Villain model\,\cite{Villain1975} for bosons of the same mass (which can be extended to bosons of different masses straightforwardly, see (25) in \cite{fil2005nondissipative}) reads
\begin{multline}
    \mathcal{S}_s(\vec{x}) = \frac{\beta}{2 m} \int_{L^2} \d \vec{x}\bigl[\tilde{n}_1 (\nabla\theta_1(\vec{x}))^2 + \tilde{n}_2 (\nabla\theta_2(\vec{x}))^2 \\
    + 2 n_\text{dr} \nabla\theta_1(\vec{x})\cdot \nabla\theta_2(\vec{x}) \bigr].
    \label{eq:Ls}
\end{multline}
In the above action, the effects of long-wavelength quantum fluctuations around the quasi-condensates have been included into the effective average densities $\tilde n_{1}$, $\tilde n_{2}$, and $n_{\textrm{dr}}$. These results only include lowest order quantum fluctuations and they are strictly valid only at low temperatures. In particular, in the vicinity of the infinite order superfluid transition, finite temperature critical fluctuations, which  are represented by vortex configurations of the phases, become relevant. These excitations are topological and need to be explicitly introduced in the action via a duality transformation in order to be treated.

\section{Vortex excitations and RG flow}
\label{sec:top_exc}
The Villain model \eqref{eq:Ls} is formulated in terms of the superfluid densities $\tilde n_i$ and $n_\text{dr}$ which already include the effect of long-wavelength phonon fluctuations. However, the phases $\theta_i \in [0,2\pi]$ are periodic and give also rise to topological vortex excitations with nonzero winding numbers \cite{berezinskii1972destruction, kosterlitz1973ordering, kosterlitz1974critical}. In this section, we derive a new RG flow equation for the renormalization of the superfluid densities due to vortex excitations. These topological excitations can be incorporated analogously to the single-component BKT case \cite{jose1977renormalization}: there is a duality transformation to a classical Coulomb gas of vortices in both components, where the nondissipative drag now introduces an interaction between them (see Appendix~\ref{sec:coulomb} for the derivation). The effective action can then be written as a sum of harmonic fluctuations and topological excitations, $\mathcal{S}_s = \mathcal{S}_\text{harm} + \mathcal{S}_\mathrm{top} + \mathcal{S}_\mathrm{core}$.  The harmonic term $\mathcal{S}_\mathrm{harm}$ resembles Eq.~\eqref{eq:Ls} but with the harmonic field $\phi(\vec{x})$ which contains no vortices. The topological term, instead, can be written as a Coulomb gas of topological charges $w_j^{(i)}$,
\begin{equation}
\begin{aligned}
    \mathcal{S}_\mathrm{top}&= -\frac{4\pi^2\beta}m \bigg ( \tilde{n}_1\sum_{j<k \in \mathcal{V}_1} C(\vec{x}^{(1)}_j - \vec{x}^{(1)}_k)w^{(1)}_j w^{(1)}_k  \\
    &\qquad +\tilde{n}_2\sum_{j<k \in \mathcal{V}_2} C(\vec{x}^{(2)}_j  - \vec{x}^{(2)}_k)w^{(2)}_j w^{(2)}_k \\
    &\qquad + n_\text{dr} \sum_{j \in \mathcal{V}_1}\sum_{k \in \mathcal{V}_2}  C(\vec{x}^{(1)}_j  - \vec{x}^{(2)}_k)w^{(1)}_j w^{(2)}_k  \biggr ), \\
    \mathcal{S}_\mathrm{core} &=  \sum_{j\in\mathcal{V}_1}  \mathcal{S}^{(1)}_{j,\text{cr}} + \sum_{j\in\mathcal{V}_2}  \mathcal{S}^{(2)}_{j,\text{cr}}.
    \label{eq:Ss}
\end{aligned}
\end{equation}
The position of the $j$th (anti-)vortex of species $i$ is denoted as $\vec{x}^{(i)}_j$, its winding number $w^{(i)}_j$, and the interaction between vortices $C(\vec{x}-\vec{y}) \equiv \ln(|\vec{x}-\vec{y}|)/2\pi$. The third line shows how the drag density gives rise to an interaction between vortices of different species. Note that also mixed vortices $\vec{x}^{(1)}_j = \vec{x}^{(2)}_k$ with winding numbers $(w^{(1)}_j, w^{(2)}_k)$ are included in this equation. However they are strongly suppressed at bare level in the case of small drag density $n_\text{dr} \ll n_{s,1},n_{s,2}$ and they cannot be thermally excited in this limit. In the intermediate drag density case one may expect these mixed vortex configurations to proliferate and introduce novel phases in the model. Nevertheless, in the weakly interacting regime relevant for present experimental realizations this should never be the case. The core contributions $\mathcal{S}_{j,\text{cr}}^{(i)}$ in the last line account for the energy cost of creating a single vortex.

The Boltzmann factor of creating a neutral vortex pair depends on their interaction energy as $p_\text{pair} \propto \exp \left [- J_{jk}C(\vec{x}_j-\vec{x}_k) \right ]$ with dimensionless coupling
\begin{equation}
    J_{jk} =\frac{4 \pi^2 \beta}m 
    \begin{cases}
        \tilde{n}_{1,2} & \text{for } j,k\in \mathcal{V}_{1,2}, \\ 
         n_\mathrm{dr} & \text{for  } j \in \mathcal{V}_1, k \in \mathcal{V}_2.
    \end{cases}
    \label{eq:J_pair}
\end{equation}
The sum over these probabilities for all neutral vortex configurations gives rise to the partition function (non-neutral configurations are strongly suppressed, see Appendix~\ref{sec:coulomb}). The interaction has to be regularized at short distance, and we use the smaller of the two scattering lengths $a \equiv \mathrm{min}[a_{11},a_{12}]$ as a short-distance cutoff such that we can use the same interaction function $C(\vec{x})$ for both components. In the low-temperature limit only vortex configurations with unit circulation $w_j^{(i)} = \pm 1$ contribute to $\mathcal{S}_\text{top}$ \footnote{Vortices with winding number $2n$ have larger energy than two vortices with winding number $n$. In the low-temperature limit, vortices of winding numbers $|w_j|>1$ are unstable with respect to the decay into vortices of smaller winding numbers~\cite{svistunov2015superfluid}. Therefore, the low-temperature phase is dominated by $w_j=\pm 1$ excitations.}. In that case, the core action is the same for all vortices within each species, $\sum_{j \in \mathcal{V}_{i}}\mathcal{S}^{(i)}_{j,\text{cr}} = 2N_d^{(i)}\mathcal{S}^{(i)}_\text{cr}$, where $N_d^{(i)}$ denotes the number of neutral vortex-antivortex dimers of species $i$. In analogy to the single-component case \cite[p.~469]{altland2010condensed} we find the topological partition function
\begin{equation}
\begin{aligned}
    \mathcal{Z}_\mathrm{top} = \sum_{N_d^{(1)},\,N_d^{(2)}=0}^{\infty}  \frac{e^{-2N_d^{(1)}\mathcal{S}^{(1)}_\text{cr}}}{(N_d^{(1)}!)^2} \times \frac{e^{-2N_d^{(2)}\mathcal{S}^{(2)}_\text{cr}}}{(N_d^{(2)}!)^2} \\
    \times \prod_{j=1}^{2N_d^{(1)}} \int_{L^2}\d^2 x_j \prod_{k=1}^{2N_d^{(2)}} \int_{L^2}\d^2 x_k \; e^{-\mathcal{S}_\mathrm{top}}.
\end{aligned}
\label{eq:Zs}
\end{equation}
The factors $(N_d!)^2$ prevent over-counting equivalent configurations in the $\sum_{jk}$ sums. In Eq.~\eqref{eq:Zs} we can interpret $\exp[-\mathcal{S}^{(i)}_\text{cr}]$ as the effective fugacity for creating a vortex of species $i$. With $\lim_{T\rightarrow 0} \mathcal{S}_\text{cr} = \infty$ we can expand the partition function in orders of $y_i \equiv e^{-\mathcal{S}^{(i)}_\text{cr}}$ as
\begin{multline}
    \mathcal{Z}_\mathrm{top} = 1 + \underbrace{y_1^2 \mathcal{Z}^{(1)}_\text{di} + y_2^2\mathcal{Z}^{(2)}_\text{di}}_{\text{dipole contributions}} \\
    + \underbrace{
    y_1^4 \mathcal{Z}^{(1)}_\text{qu} + y_1^2 y_2^2 \mathcal{Z}^{(1)(2)}_\text{qu} +  y_2^4 \mathcal{Z}^{(2)}_\text{qu}}_{\text{quadrupole contributions}} + \mathcal{O}(y_{1,2}^6).
   \label{eq:Zs2}
\end{multline}
At zero temperature $y_i\to0$ and no unbound vortices are present; at finite temperature, the vortex density is controlled by the respective Boltzmann factors $y_{1,2}$. 

\begin{figure}[t]
    \centering
    \includegraphics[width=1.0\linewidth]{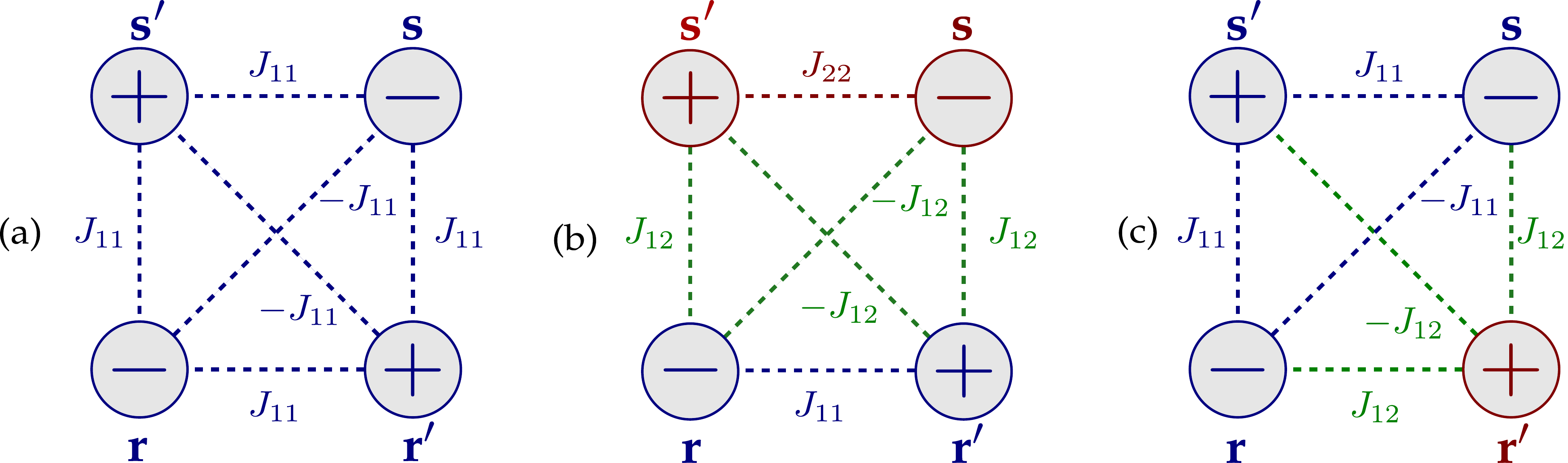}
    \caption{Illustration of the dipole configurations ($\vec{s},\vec{s}'$) of order $y^2$ with two test charges ($\vec{r},\vec{r}'$) which appear in Eq.~\eqref{eq:Zs2}. The blue and red colors refer to vortex excitations in the first and second species, respectively. Their action is calculated below.}
\label{fig:diagrams_y1}
\end{figure}
In analogy to the single-component case, vortices disrupt the superfluid flow. In order to determine the renormalization of the superfluid densities, one needs to evaluate the interactions between vortices of both species. The effective probability of creating a vortex pair is then given by the expectation value $p_\text{pair}^{\text{eff}} = \langle e^{-J_{ij}C_{\vec{rr'}}}\rangle$, which includes thermal excitation of additional dipoles according to Eq.~\eqref{eq:Zs2}. Without loss of generality, we can choose a negative test charge (winding number) $\ominus$ at $\vec{r}$ and a positive charge $\oplus$ at $\vec{r}'$. We want to compute the effect on these test charges by a dipole within the thermal ensemble, with charge $\ominus$ at $\vec{s}$ and $\oplus$ at $\vec{s}'$, see Fig.~\ref{fig:diagrams_y1} for illustration. The contributions from these three exemplary configurations can be written as follows,
\begin{enumerate}[(a)]
    \item $J_{11}(C_{\vec{ss'}}+C_{\vec{rr'}}-(C_{\vec{rs}}+C_{\vec{r's'}}-C_{\vec{r's}}-C_{\vec{rs'}})) \\ 
    = J_{11}(C_{\vec{ss'}}+C_{\vec{rr'}} - D_{\vec{rr'ss'}})$
    \item $J_{22}C_{\vec{ss'}}+ J_{11}C_{\vec{rr'}}-J_{12}(C_{\vec{rs}}+C_{\vec{r's'}}-C_{\vec{r's}}-C_{\vec{rs'}}) \\
    = J_{22}C_{\vec{ss'}}+ J_{11}C_{\vec{rr'}}-J_{12}D_{\vec{rr'ss'}}$
    \item $J_{11} (C_{\vec{ss'}}-C_{\vec{rs}}+C_{\vec{rs'}}) + J_{12}( - C_{\vec{r's'}}+C_{\vec{rr'}}+C_{\vec{r's}})$,
\end{enumerate}
where we abbreviated $C(\vec{x} - \vec{x'})$ by $C_{\vec{x}\vec{x'}}$ and defined the dipole moment $D_\vec{rr'ss'} = C_{\vec{rs}}+C_{\vec{r's'}}-C_{\vec{r's}}-C_{\vec{rs'}}$. The signs arise from the combination of winding numbers which multiply the coupling. We can decompose the effective probability $p_\text{pair}^{\text{eff}}$ into terms where only the same or the opposite species appears (non-mixed), as in (a) and (b), and in terms with mixed contributions as in (c). These contributions give rise to a screening of the bare interaction, which can be incorporated into a renormalization of the superfluid densities. We find an extended set of RG flow equations to leading order in fugacity (see Appendix~\ref{sec:RG_flow} for the derivation), for an increasing spatial length scale $r\simeq a e^l$,
\begin{align}
    \frac{\d \tilde{n}_{\ez}^{-1}}{\d l} &=  \frac{4\pi^3 \beta}{m}\left(y_\ez^2 + {y^2_\ze\left(\frac{ n_\mathrm{dr}}{\tilde{n}_\ez} \right)^2} \color{black} \right) + \mathcal{O}(y_{1,2}^4),\notag\\
    \quad \frac{\d y_\ez}{\d l} &= \left(2 -  \frac{\pi \beta }{m}\tilde{n}_\ez \right)y_\ez + \mathcal{O}(y_{1,2}^3),\label{eq:RG}\\
    \frac{\d n^{-1}_\text{dr}}{\d l} &= \frac{4\pi^3\beta}{m} \, n_\text{dr}^{-1}\left ( y_1^2 \tilde{n}_1 + y_2^2 \tilde{n}_2 \right  ) + \mathcal{O}(y_{1,2}^4)  \notag.
\end{align}
The initial conditions for $\tilde{n}_i$ and $n_\mathrm{dr}$ are given by the coefficients in the Villain model \eqref{eq:Ls}, which are computed from Eqs.~\eqref{eq:rho_ni} and \eqref{eq:drag} in Appendix~\ref{sec:andreev}. According to the conventional BKT argument\,\cite{jose1977renormalization}, the microscopic (bare) vortex fugacity is given by the formula $y_i= e^{-\pi^2\beta \tilde{n}_i/2m}$ as in the single layer XY model. Further corrections arising from the interaction between the components are negligible at the present expansion order. The RG equations\,\eqref{eq:RG}  can then be integrated from the microscopic scale $l=0$ up to the physical scale $l$.

\section{Results}
\label{sec:results}
\begin{figure*}[t!]
	\centering
	\subfigure[$\quad$]{\label{Fig6a}\includegraphics[width=.32\textwidth]{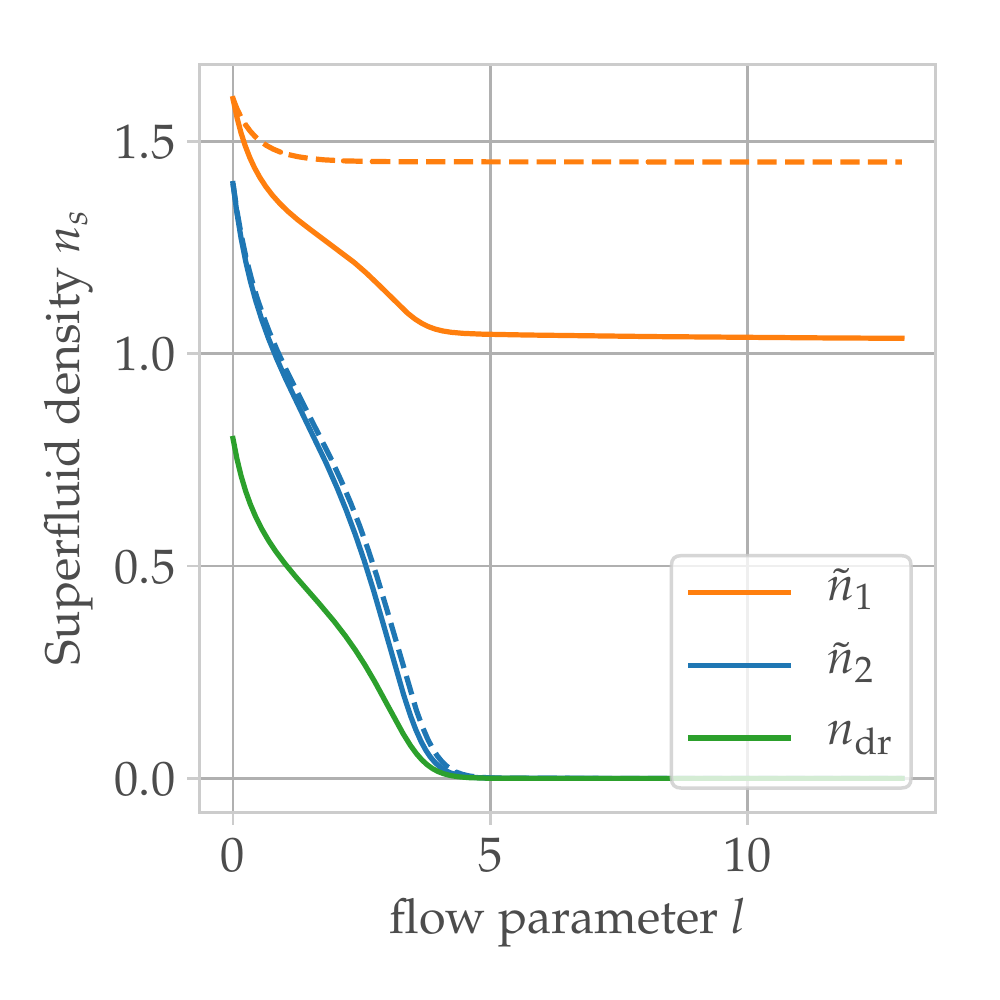}}
	\hfill
	\subfigure[$\quad$]{\label{Fig6b}\includegraphics[width=.32\textwidth]{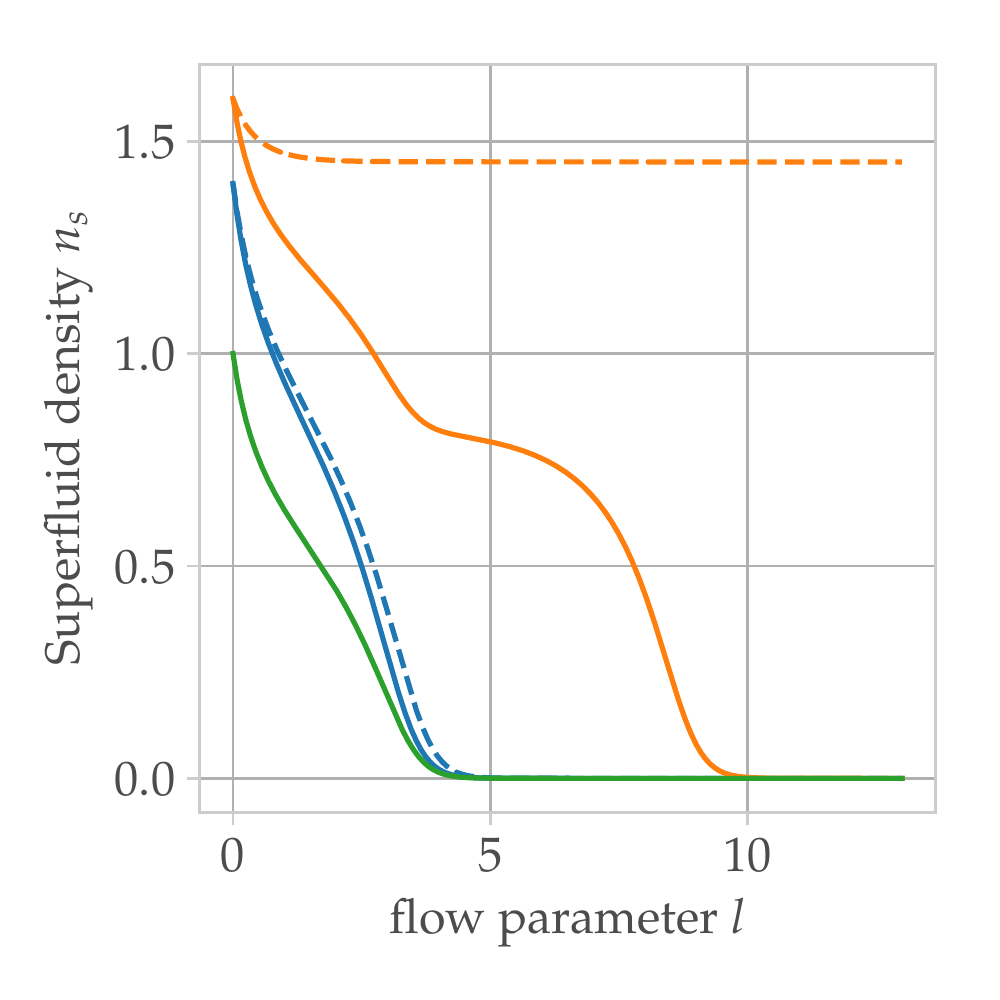}}
	\hfill
	\subfigure[$\quad$]{\label{Fig6c}\includegraphics[width=.32\textwidth]{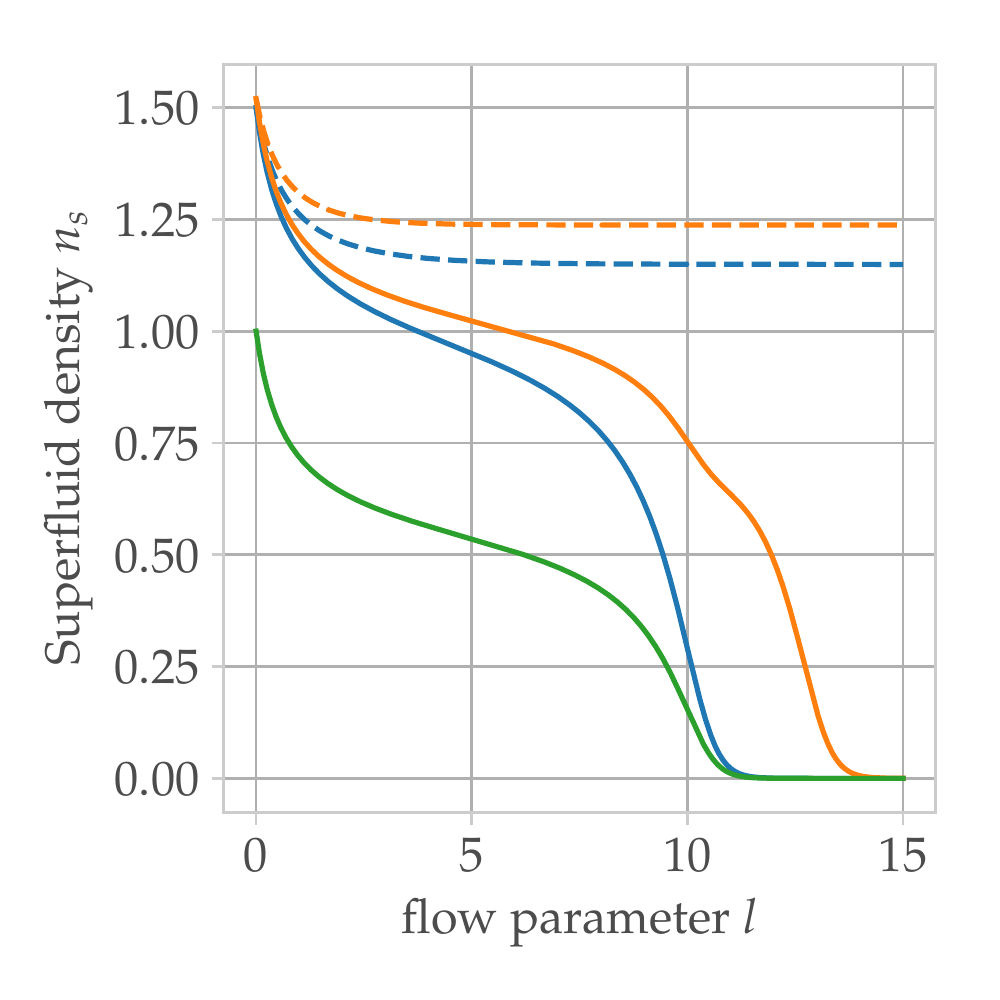}}
	\caption{\label{Fig6}Exemplary BKT flow trajectories. The dashed lines illustrate the conventional flow of the uncoupled system, while solid curves represent the coupled flows obtained for the BKT system in Eq.\,\eqref{eq:RG}. Apart from the case of irrelevant vortex configurations where both components remain superfluid, three major cases arise: if one of the bare superfluid densities is smaller than the uncoupled critical value [middle curves in panels (a) and (b)], already small drag densities may induce strong renormalization of the superfluid density in the majority component, see panel (a). At larger drag densities, this renormalization may become so strong as to induce vortex unbinding also in the majority component, which would have remained superfluid in the uncoupled case, see panel (b). Finally, a coupled breakdown of superfluidity appears for low enough bare superfluid densities $\bar{n}_{1}\approx\bar{n}_{2}\gtrsim n_{c}^{0}$, where both components were superfluid in the uncoupled case (dashed lines) but drag induced fluctuations lead the system to the normal phase, see figure (c).}
\end{figure*}

Our flow equations \eqref{eq:RG} extend the traditional BKT equations for uncoupled superfluids by new terms proportional to $n_\text{dr}$, which introduce a coupling between both superfluid densities during the RG flow. The new RG equations quantify how vortices in one component influence vortices in the other, and have the tendency to suppress superfluidity. For the uncoupled system with vanishing drag $n_\text{dr}=0$, we recover two separate single-component BKT flows for $n_{s,i}$ and $y_i$. In this uncoupled case, there is a single critical superfluid density $n_{c}^{0}(T)$ at temperature $T$; if either one of the two bare superfluid densities $\bar{n}_{i,b}$ is below this critical value, vortex unbinding occurs in that component and drives the renormalized superfluid density to zero, while the respective vortex fugacity diverges, $y_{i}\to\infty$. In the following, we will describe how a finite coupling between the two components, induced by the drag density, modifies this picture.

Assume that without drag $n_\mathrm{dr}=0$ the superfluid density in the first component is renormalized to some finite value $\bar{n}_{1,b}>n_{c}^{0}$, while the second component renormalizes to zero $\bar{n}_{2,b}<n_{c}^{0}$, as illustrated in Fig.~\ref{Fig6a}; this corresponds to a flow with $y_1\rightarrow 0$ and $y_2\rightarrow \infty$. Instead, when the drag $n_\text{dr}\neq0$ is included, $\tilde{n}_{1}$ will continue to decrease beyond the fixed point of the uncoupled case. This can be understood analytically in the limit $y_1\to0$, where the flow equation \eqref{eq:RG} for $\tilde{n}_1$ reads
\begin{equation}
     \frac{\d \tilde{n}_{1}^{-1}}{\d l} =  \frac{4\pi^3 \beta}{m}y^2_2\left(\frac{n_\mathrm{dr}}{\tilde{n}_1} \right)^2 .
     \label{eq:n1flow}
\end{equation}
Hence, if one of the two superfluid densities is renormalized to zero, it drags the other one to a lower density as well. Moreover, for large enough drag densities, such additional renormalization also drives the first component to the normal state, completely disrupting superfluidity, see Fig.\,\ref{Fig6b}. Along the same line, we also find a \emph{coupled superfluidity breakdown} regime, illustrated in Fig.\,\ref{Fig6c}: while the uncoupled case would have finite superfluid densities in both components, a large enough drag density renormalizes both of them to zero. 

The observation of coupled superfluidity breakdown is expected for values of the drag density $n_{\text{dr}}$ comparable to the ones of the single components depleted densities. Specifically, for 
a $^{87}$Rb mixture in two hyperfine states with $a_{12}\approx a_{11}\approx 100a_{B}$
the largest $n_{\text{dr}}$ is reached for equal densities at a temperature $T\approx 2\mu K$.
In this configuration one has $n_{\text{dr}}\approx 10\mu m^{-2}$, see Fig.\,\ref{fig:drag_densities}, which is very close
to the depleted densities for the same parameters, see Fig.\,\ref{fig:superfluid_densities}.

In the uncoupled case the critical temperature is given by the Kosterlitz-Nelson criterion $\tilde{T}_c^{(i)} = \pi \hbar^2 \tilde{n}_i/2 k_B m$ in terms of the renormalized superfluid density $\tilde n_i$ at the end of the RG flow. In the coupled case, it follows from the RG equations \eqref{eq:RG} that there are two critical temperatures in the coexistence regime which differ from the uncoupled case. According to this model, the locking of superfluidity will occur for $n_\text{dr}$ large enough compared to $n_1$ and $n_2$. Note that it is not possible to have a finite drag $n_\text{dr}$ in the high-temperature phase where $\tilde{n}_{1,2} \to 0$: if $n_\text{dr} < \tilde n_{1,2}$ initially, then it will always remain smaller by the flow equation, such that $n_\text{dr}$ decreases to zero as well.

\begin{figure}[t]
    \centering
    \includegraphics[width=1.0\linewidth]{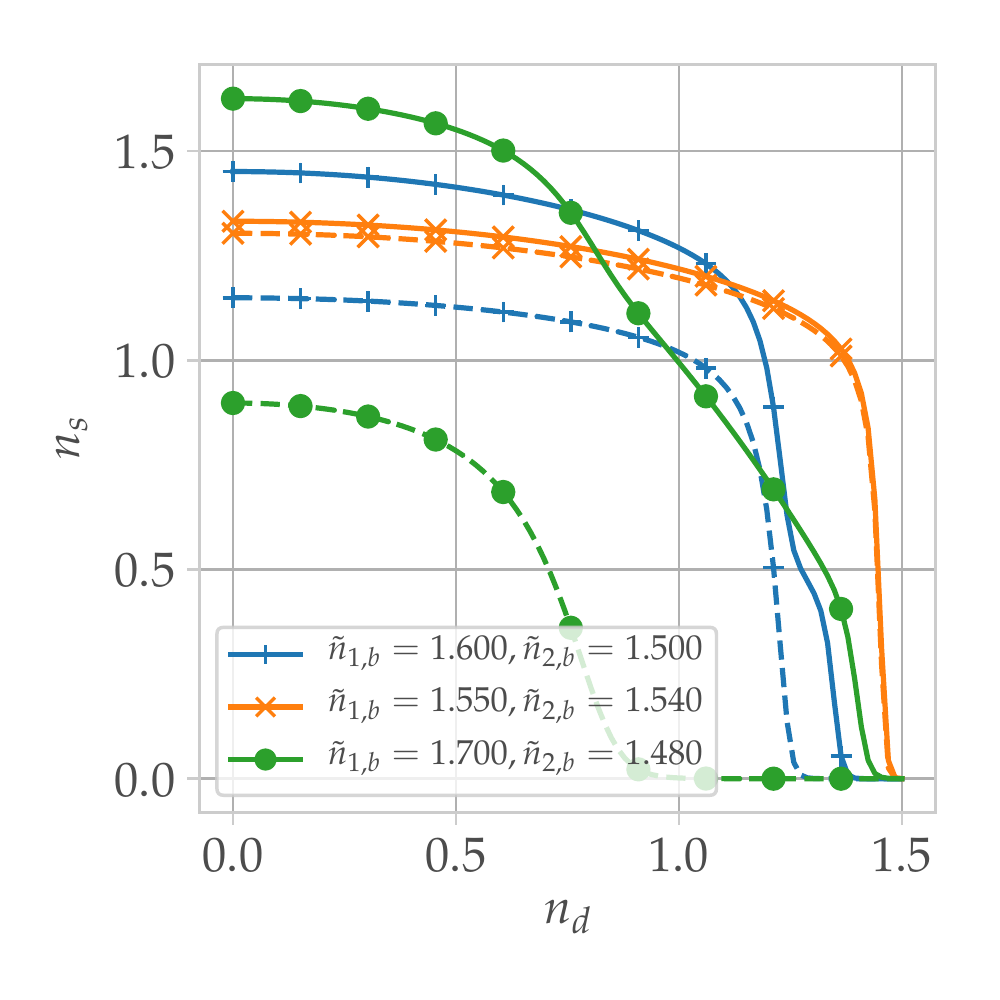}
    \caption{Renormalized superfluid densities as a function of the initial drag density, for various bare initial conditions. The majority superfluid density $\bar{n}_{1}$ is shown as a solid line, while $\bar{n}_{2}$ is represented by a dashed line. When the bare values for the two superfluid components are well separated, also the renormalized values always remain apart, even at large initial drag densities, see the blue and green curves. Nevertheless, when the bare superfluid densities $n_{1,b}$ and $n_{2,b}$ are close enough, the locking mechanism brings them closer and finally merges them at large drag densities, see the orange lines.}
\label{Fig7}
\end{figure}

Our results suggest that finite drag densities $n_\text{dr}$ may couple the two superfluids so strongly that the collapse of one component can lead to the collapse of the other or, even more surprisingly, two stable superfluids in the uncoupled regime can be driven above criticality and disappear if a strong enough coupling is introduced. Therefore, the finite drag density can introduce a locking effect of the two critical temperatures, which tend to become equal in the intermediate coupling limit. Such an effect only appears for close enough superfluid densities in the two components, as shown in Fig\,\ref{Fig7}.

\section{Conclusion}
\label{sec:conclusion}
We find that vortex excitations in a binary Bose mixture with the same mass~\footnote{However, since \eqref{eq:Ls} can be extended straightforwardly to the case of different masses, the RG flow is valid also in that case.}  give rise to a coupled breakdown of superfluidity: whenever one of the superfluids surpasses the single-component critical temperature and vanishes, it can lead to the collapse of the other component as well, given the drag density $n_\text{dr}$ between them is large enough. In that case, they share a unique critical temperature. This result is qualitatively different from the uncoupled case, where each component can have a different critical temperature depending on its density in the coexistence regime.  We thus observe how topological excitations of the phase of one component have a large influence on the superfluid properties of the other. 

Our  derivation is perturbative in the fugacity and is in principle valid only for small $y_1$ and $y_2$. Nevertheless, in analogy with the traditional BKT case, we expect the flow equations \eqref{eq:RG} to capture all universal aspects of the system, at least as long as mixed vortex configurations do not proliferate~\footnote{Mixed vortex contributions turn up only at fourth order due to charge neutrality, see Appendix D.}.
In order to include these, one should use a theoretical approach which incorporates both density and phase fluctuations (including topological excitations) nonperturbatively, which can be achieved, e.g., using the functional renormalization group~\cite{defenu2017nonperturbative}. Such an accurate treatment of density fluctuations is especially important near the quantum phase transition where fluctuations of the relative density become soft. Nevertheless, our flow equations already show that in the coexistence regime not too close to the phase boundary, the superfluid transition temperatures are locked, a new effect which is not observed in the uncoupled case and inaccessible in mean-field calculations.

It is important to note that this analysis is obtained for the case $g_{11},g_{12}>0$, but should in principle be extendable to the more general case. However, since in the limit $g= \frac{1}{2}(g_{11}+g_{12})=0$ the speed of sound of the lower branch becomes zero, in that case no superfluid is to be expected. Therefore, in order to see superfluid behavior, the regime $g > 0$ is appropriate. The breakdown of superfluidity is most striking in the regime where one component is superfluid while the other is normal, $\tilde T_c^{(1)} < T < \tilde T_c^{(2)}$, which could be achieved by fine-tuning the difference of the chemical potentials $\Delta \mu \neq 0$. In that case, the superfluid behavior in the majority component is disrupted by proliferating vortices in the minority component, and will eventually collapse for strong enough drag. An experimental test of our predictions appears viable with present technology for ultracold binary Bose mixtures~\cite{cabrera2018quantum, ye2018dressed, schulze2018feshbach}. 

We acknowledge stimulating discussions with Thomas Gasenzer, Johannes Hofmann, Andrea Trombettoni, and in particular Gergely Zar\'and, who proposed the problem. This work is supported by Deutsche
Forschungsgemeinschaft (DFG) via Collaborative Research Centre SFB 1225 (ISOQUANT) and under Germany’s Excellence Strategy EXC-2181/1-390900948 (Heidelberg STRUCTURES Excellence Cluster).

\appendix
\section{Low-temperature excitations: normal modes and quantum fluctuations}
\label{sec:phonon}
The complex fields $\psi_i(x,t) = \sqrt{\bar{n}_i} + \delta \psi_i(x,t)$ fluctuate around the 
quasi-condensates $\bar{n}_1,\bar{n}_2$ such that $\left \langle \psi_i(x,t) \right \rangle = \sqrt{\bar{n}_i}$ and $\left \langle \delta \psi_i(x,t)\right \rangle = 0$. The Lagrangian for the fluctuation independent term is given by $\mathcal{L}_\mathrm{pot}[\bar{n}_1,\bar{n}_2]$ in Eq.~\eqref{eq:Lpot}, whereas the fluctuating part can be written as
\begin{equation}
    \mathcal{L}_\mathrm{fl} = \sum_i \, \, \, \mathcal L^{(1)}_i +  \mathcal L^{(2)}_i + \mathcal L^{(3)}_i + \mathcal L^{(4)}_i,
\end{equation}
where $\mathcal L_i^{(1)}$ for component $i$ consists of terms linear in $\delta \psi$, $\mathcal L_i^{(2)}$ of quadratic terms, etc.  These are given by
\begin{equation}
    \begin{aligned}
        \label{eq:fluctuation_expansion}
        \mathcal{L}^{(1)}_1 &= \delta \psi^\dagger_1(-\mu_1 + g_{11}\bar{n}_1 + g_{12}\bar{n}_2)\sqrt{\bar{n}_1} + \mathrm{h.c.}, \\
        \mathcal{L}^{(2)}_1 &= \delta \psi_1^\dagger (\partial_\tau - \tfrac{\nabla^2}{2m} - \mu_1 + 2 g_{11}\bar{n}_1 + g_{12}\bar{n}_2)\delta \psi_1 \\ & \quad + \tfrac{1}{2}g_{11}\bar{n}_1 ((\delta \psi^\dagger_1)^2 + \delta \psi_1^2)  \\
        & \quad +\tfrac{1}{2}g_{12} \sqrt{\bar{n}_1 \bar{n}_2}(\delta \psi^\dagger_1 \delta \psi^\dagger_2 + \delta \psi^\dagger_1 \delta \psi_2 + \mathrm{h.c.}), \\
        \mathcal L^{(3)}_1  &=  g_{11}\sqrt{\bar{n}_1}(\delta \psi_1^\dagger)^2\delta \psi_1 + g_{12}\sqrt{\bar{n}_1}\delta \psi^\dagger_2 \delta \psi_2 \delta \psi_1 + \mathrm{h.c.}, \\
        \mathcal L^{(4)}_1 &= \tfrac{1}{2}g_{11}(\delta \psi^\dagger_1 \delta \psi_1)^2 + \tfrac{1}{2}g_{12}(\delta \psi^\dagger_1 \delta \psi^\dagger_2 \delta \psi_1\delta \psi_2),
    \end{aligned}
\end{equation}
and similar terms arise for the second component $\mathcal L_2^{(j)}$. At this point, we neglect higher terms than quadratic ones, insert the mean-field densities \eqref{eq:density_MF} to eliminate $\mathcal{L}_i^{(1)}$ and perform the Fourier transform. The Lagrangian can then be written as
$\mathcal L_\text{fl} = \frac{1}{2}\,\delta \psi^\dagger\, \mathcal{M}\,\delta \psi $ with 
$\delta \psi^\dagger = (\delta \psi^\dagger_1,\delta \psi_1,\delta \psi^\dagger_2,\delta \psi_2)$ and the quadratic form
\begin{widetext}
\begin{equation}
  \mathcal{M} = \begin{pmatrix}\epsilon_{k} + g_{11} \bar n_{1} - \omega & g_{11} \bar n_{1} & g_{12} \sqrt{\bar n_{1} \bar n_{2}} & g_{12} \sqrt{\bar n_{1} \bar n_{2}}\\
  g_{11} \bar n_{1} & \epsilon_{k} + g_{11} \bar n_{1} + \omega & g_{12} \sqrt{\bar n_{1} \bar n_{2}} & g_{12} \sqrt{\bar n_{1} \bar n_{2}}\\
  g_{12} \sqrt{\bar n_{1} \bar n_{2}} & g_{12} \sqrt{\bar n_{1} \bar n_{2}} & \epsilon_{k} + g_{11} \bar n_{2} - \omega & g_{11} \bar n_{2}\\
  g_{12} \sqrt{\bar n_{1} \bar n_{2}} & g_{12} \sqrt{\bar n_{1} \bar n_{2}} & g_{11} \bar n_{2} & \epsilon_{k} + g_{11} \bar n_{2} + \omega\end{pmatrix}.
\end{equation}
\end{widetext}
The dispersion relation $\omega_{k\pm}$ is found by solving $\det(\mathcal{M})=0$, with the result \eqref{eq:phonon_freq} given in the main text. These normal modes are composite modes consisting of excitations in both species. One can identify the $\pm$ mode with in-phase (density) and out-of-phase (spin) variations with respect to the two species~\cite{larsen1963binary}. Since the frequencies are gapless, we can identify them with the two phonon modes of the system, which are expected by the Goldstone theorem~\cite{weinberg1995quantum}. Like in the single-component case, the Matsubara sum of the action can be evaluated using convergence-factor regularization \cite{altland2010condensed}. The grand potential of the fluctuating part $\Omega_\mathrm{fl} = -T \ln \mathcal{Z}_\mathrm{fl}$ is given by
\begin{equation}
\begin{aligned}
    \Omega_\mathrm{fl} =  \sum_{\vec k\sigma} \left( \frac{1}{2}\left( \omega_{k\sigma} - \epsilon_k - m c^2_\sigma \right) + T \ln ( 1 - e^{-\beta \omega_{k \sigma}}) \right).
    \label{eq:omegafl}
\end{aligned}
\end{equation}
The first term is not temperature dependent and gives rise to quantum fluctuations
\begin{equation}
\begin{aligned}
    \Omega_\mathrm{qfl}
    &= \frac 12\sum_{\vec k\sigma} \left(\omega_{k\sigma} - \epsilon_k - m c^2_\sigma \right)\\
    &= \frac{V m}{4\pi} \sum_\sigma\int_0^{\epsilon_0} \d \epsilon \left(\sqrt{\epsilon(\epsilon + 2m c^2_\sigma)} - \epsilon - m c^2_\sigma \right).
\end{aligned}
\label{eq:omega_fl}
\end{equation}
The integral is formally ultraviolet (UV) divergent and needs to be regularized by a UV cutoff scale $\epsilon_0\gg m c_\pm^2$, as in the single-component case \cite{salasnich2016zero}. The pressure due to quantum fluctuations then takes the form \cite{petrov2016ultradilute}
\begin{equation}
    \mathrm{p}_\mathrm{fl}(T=0) = -\frac{m}{8\pi} \sum_\sigma m^2c^4_\sigma \left (\ln \left(\frac{m c^2_\sigma}{2 \epsilon_0}\right) + \frac{1}{2} \right).
    \label{eq:pfl}
\end{equation}
A priori, different regularization scales $\epsilon_\pm$ could be chosen for the two branches. However, as the fluctuation pressure depends only logarithmically on these cutoff scales, one can choose a common scale $\epsilon_0$ for both, up to subleading logarithmic corrections $\ln(\epsilon_\pm/\epsilon_0)$ which are small \cite{petrov2016ultradilute}. Hence, the quantum fluctuation pressure can be reabsorbed into the expression for the mean-field pressure $p_0 = \mu^2/2g(\epsilon_0) = (m\mu^2/8\pi) \ln(\epsilon_b/\epsilon_0)$ by a redefinition of the coupling. Whereas the mean-field pressure is defined in terms of the bare coupling $g(\epsilon_0)$ from Eq.~\eqref{eq:gE12} evaluated at the cutoff energy, the additional quantum fluctuation part \eqref{eq:pfl} effectively shifts the regularization scale from the UV scale $\epsilon_0$ to the many-body scale $\mu \sim m c_\pm^2$, and we find $p(T=0)=\sum_\sigma m^2c_\sigma^4/2g(\mu)$, up to logarithmic corrections.

\section{Two-component superfluidity and the Andreev-Bashkin effect}
\label{sec:andreev}
In this section we give the expressions for the two-component normal fluid densities and the drag density, following the derivation given in~\cite{fil2005nondissipative, Konietin2018}. Both normal densities arise only from thermal fluctuations and vanish at zero temperature. The normal fluid densities are given in terms of the normal mode frequencies $\omega_{k\pm}$ as
\begin{equation}
\begin{aligned}
   n_{n,i} &= - \frac{1}{2L^2} \sum_{\vec{k}} \epsilon_k \bigg[ \frac{\d n(\omega_{k+})}{\d \omega_{k+}} \left(1 \pm \frac{\omega^2_{k1} - \omega^2_{k2}}{\omega^2_{k+} - \omega^2_{k-}}\right ) \\
   &\qquad \qquad \quad+ \frac{\d n(\omega_{k-})}{\d \omega_{k-}} \left(1 \mp \frac{\omega^2_{k1} - \omega^2_{k2}}{\omega^2_{k+} - \omega^2_{k-}} \right ) \bigg] \\
   &= - \frac{1}{2L^2} \sum_{\vec{k}\sigma} \epsilon_k \frac{\d n(\omega_{k\sigma})}{\d \omega_{k\sigma}} \left(1 \pm \sigma  \gamma \right )
\end{aligned}
\label{eq:rho_ni}
\end{equation}
with sign $\pm$ for component $i=1,2$, and $\gamma = (\omega^2_{k1} - \omega^2_{k2})/(\omega^2_{k+} - \omega^2_{k-}) = g_{11}\Delta \bar n/(mc^2_- - mc^2_+)$ independent of $k$, while $n(\omega)=(e^{\beta \omega} - 1)^{-1}$ denotes the bosonic occupation number. The drag density, in turn, is given by
\begin{equation}
\begin{aligned}
   n_\text{dr}  &= 2 L^{-2}\sum_\vec{k} \frac{g_{12}^2 \bar n_1 \bar n_2 \epsilon_k^3}{\omega_{k+} \omega_{k-}} \biggl[ \frac{1 + n(\omega_{k+}) + n(\omega_{k-})}{(\omega_{k+} + \omega_{k-})^3} \\
   &\qquad- \frac{n(\omega_{k+}) - n(\omega_{k-})}{(\omega_{k+} - \omega_{k-})^3} \\
   &\qquad+\frac{\omega_{k+} \omega_{k-}}{(\omega_{k+}^2 - \omega_{k-}^2)^2}\left ( \frac{\d n(\omega_{k+})}{\d \omega_{k+}} + \frac{\d n(\omega_{k-})}{\d \omega_{k-}}\right )\biggr].
\end{aligned}
\label{eq:drag}
\end{equation}
In the zero-temperature limit, the normal densities vanish while the drag density reaches the finite value due to quantum fluctuations \cite{Konietin2018},
\begin{equation}
    n_\text{dr}(T=0)
    = \frac{g_{12}^2 \bar n_1 \bar n_2}{8\pi}\,
    \frac{c_+^4 - c_-^4 - 4c_+^2 c_-^2 \ln(c_+ / c_-)} {(c_+^2 - c_-^2)^3}.
\end{equation}
The drag densities are plotted in Fig.~\ref{fig:drag_densities}: they diverge at the phase transition where the normal modes $\omega_{k\pm}$ become soft. The meaning of the drag density $n_\text{dr}$ becomes apparent when computing the supercurrents from the fluctuation grand potential $\Omega_\text{fl}(\vec v_i)$ at finite temperature $T>0$ and superflow $\vec v_i$. To linear order in $\vec v_i$, the supercurrents $\vec j_i$ are then determined by the normal and drag densities as given in Eq.~\eqref{eq:density_s} of the main text.

The nondissipative drag thus changes the qualitative behavior in the coexistence regime: for instance, a superflow only in component 1 with $\vec v_1\neq0$ and $\vec v_2=0$ nevertheless yields a supercurrent $\vec j_2=n_\text{dr}\vec v_1\neq0$ in the same direction also in the second component \cite{andreev1975af, bashkin1997instability, svistunov2015superfluid}.

\section{Derivation of the two-component vortex-antivortex Coulomb gas}
\label{sec:coulomb}
In this appendix we derive the vortex-antivortex Coulomb gas, which is dual to the
phase fluctuations of the two-component Bose gas. Starting with the two-component Villain model \eqref{eq:Ls}, we can decompose $\nabla \theta_i$ into a curl-free part $\nabla \phi_i$ and a divergence-free part $\vec A_i$,
\begin{equation}
    \nabla \theta_i(\vec{x})
    = -\nabla \phi_i(\vec{x}) + \nabla \times \vec{A}_i(\vec{x}).
    \label{eq:helmholtz}
\end{equation}
 In two dimensions, $\nabla \times \vec{A}_i(\vec{x}) = \nabla \times (\vec{e}_z \chi_i(\vec{x}))$ with a scalar function $\chi_i(\vec{x})$. Also, $e^{i\theta_i}$ must be single-valued and thereby gives rise to an integer winding number $w^{(i)}\in \mathbb{Z}$ defined as
\begin{equation}
\begin{aligned}
    2\pi w^{(i)} &= \oint_C \nabla \theta_i \cdot \d \vec\ell = \int_{L^2} \d^2 x [\nabla \times \nabla \theta_i] \\
    &= \int_{L^2}\d^2 x  \left[ \nabla \times (\nabla \times \vec{A}_i(\vec{x})) -\nabla \times (\nabla \phi_i(\vec{x})) \right] \\
    &= \int_{L^{2}}\d^2 x \nabla^2 \chi_i(\vec{x}) .
\end{aligned}
\label{eq:2pin}
\end{equation}
The scalar function $\phi_i$ is the harmonic part of $\theta_i$ without vortex excitations. To leading order, the harmonic contributions $\phi_i$ and topological excitations $\chi_i$ decouple~\cite{altland2010condensed}, i.e., $\mathcal{S}_s =\mathcal{S}_\mathrm{harm} +\mathcal{S}_\mathrm{top}$. Let us therefore focus on $\mathcal{S}_\mathrm{top}$, which includes the nontrivial topological solutions of the Poisson equation in two dimensions. Since the fundamental group $\pi_1(S_1) \cong \mathbb{Z}$, we can decompose $w^{(i)} = \sum_{j\in\mathcal{V}_i} w_j^{(i)}$,
where $w^{(i)}_j \in \mathbb{Z}$ is the topological charge of the $j$th vortex within the $i$th component, with $j\in \mathcal{V}_i = \{1,...,N_i\}$ and $N_i$ vortices in total. Eq.~\eqref{eq:2pin} can be rephrased in terms of vortices centered at position $\vec{x}_j$ as
\begin{equation}
\begin{aligned}
   \nabla^2 \chi_i(\vec{x}) &= 2\pi \sum_{j \in \mathcal{V}} w^{(i)}_j \, \delta^2(\vec{x} - \vec{x}_j) \\
   \Rightarrow \chi_i(\vec{x}) &= \sum_{j \in \mathcal{V}} w^{(i)}_j \, \ln(\abs{\vec{x} - \vec{x}_j}),
   \label{eq:chi}
\end{aligned}
\end{equation}
where $\chi_i(\vec x)$ is the solution of the two-dimensional inhomogeneous Laplace equation. The action for topological excitations can be integrated by parts as
\begin{multline}
    \int_{L^2} \d^2 x \, (\nabla \theta_i)^2
    = \int_{L^2} \d^2 x \,(\nabla \times \vec{e}_z \chi_i(\vec{x}))^2 \\= [\chi_i(\vec{x}) \nabla \chi_i(\vec{x})]_{\partial L^2} - \int_{L^2} \d^2 x \, \chi_i(\vec{x})\nabla^2\chi_i(\vec{x}),
\end{multline}

where the first contribution is evaluated at the boundary $\partial L^2$ of the integration area. The boundary term vanishes for $\sum_{j \in \mathcal{V_i}} w_j^{(i)} = 0$, i.e., topological charge neutrality, and diverges otherwise. Let us therefore assume charge neutrality, since non-neutral configurations are suppressed strongly. We can then use the identity \eqref{eq:chi} and find
\begin{equation}
\begin{aligned}
     \int_{L^2} \d^2 x \, (\nabla \theta_i)^2
     &= - 2\pi\sum_{j,k \in \mathcal{V}_i} \, w^{(i)}_j w^{(i)}_k \ln(\abs{\vec{x}_j - \vec{x}_k}) \\
     &\equiv - 4\pi^2\sum_{j,k \in \mathcal{V}_i} \, w^{(i)}_j w^{(i)}_k C(\vec{x}_j - \vec{x}_k)
     \label{eq:v2}
     \end{aligned}
\end{equation}
with interaction $C(\vec{x}-\vec{y}) \equiv \ln(|\vec{x}-\vec{y}|)/2\pi$. At this point a divergences arises (i) for terms $j=k$ in the sum and (ii) for vortex configurations with $\vec{x}_j = \vec{x}_k$ for $j\neq k$. While (i) can be cured easily by assuming a small, but finite self-interaction, (ii) arises due to the failure of the continuous description for very small length scales. Let us therefore assume the action to be finite. 

Finally, the third contribution in Eq.~\eqref{eq:Ls} has the form $\nabla \theta_1(x)\cdot\nabla \theta_2(x)$, and in analogy to \eqref{eq:v2} we arrive at
\begin{multline}
 \int_{L^2} \d^2 x \, \nabla \theta_{1}(x)\cdot \nabla \theta_{2}(x) \\
 = -4\pi^2 \sum_{j \in \mathcal{V}_1}\sum_{k \in \mathcal{V}_2} w^{(1)}_j w^{(2)}_k C(\vec{x}^{(1)}_j - \vec{x}^{(2)}_k),
\end{multline}
where $w^{(i)}_j$ and $\vec{x}^{(i)}_j$ denote the topological charge and the position of the $j$th Vortex of species $i$ (the two sets $\mathcal{V}_1,\mathcal{V}_2$ are in general different). The singularities are of type (ii) and will be cured analogously to make all contributions of the action finite, and we derive \eqref{eq:Ss} of the main text.

The contributions of self-interaction type (i) in Eq.~\eqref{eq:v2} with $j=k$ have been grouped into the core action $\mathcal{S}_\text{cr}$, and the sum restricted to different vortices $\sum_{j\neq k}$. The core action defines the energy needed to excite a single vortex, i.e., the chemical potential of a vortex; it is finite but depends on the short-distance details of the system. A popular choice is to use $\mathcal S_\text{cr} = \pi^2\beta \tilde n_i/2m$ of the lattice XY model.

\section{Derivation of the RG flow}
\label{sec:RG_flow}
In this appendix we derive the RG flow equations by following standard procedure \cite{altland2010condensed, svistunov2015superfluid}, and then showing what changes for two components. The flow equations arise from the following argument: Two test charges $\ominus$ at $\vec r$ and $\oplus$ at $\vec r'$, say both of component 1, have a direct interaction $J_{11}C_{\vec r\vec r'}$. In addition, the interaction with thermally excited charges $\vec s$, $\vec s'$ leads to an induced interaction between $\vec r$ and $\vec r'$; this can be incorporated into a renormalization of $J_{11}$. The induced interaction is found to leading order $\mathcal O(y_{1,2}^2)$ by considering vortex configurations of the type shown in Fig.~\ref{fig:diagrams_y1}. Specifically for test charges $\vec r$, $\vec r'$ both of component 1, there are two contributions: either $\vec s$, $\vec s'$ are also of component 1 (Fig.~\ref{fig:diagrams_y1}a), or both $\vec s$, $\vec s'$ are component 2 (Fig.~\ref{fig:diagrams_y1}b). Mixed configurations with $\vec s$ of component 1 and $\vec s'$ of component 2 are suppressed because they violate charge neutrality; mixed-species dipoles appear only at order $\mathcal O(y_{1,2}^4)$. Throughout our phase diagram the drag density remains small, $n_\text{dr} \ll \tilde n_{1,2}$, so in our case the quadratic approximation is justified. However, in Bose droplets \cite{petrov2015quantum, petrov2016ultradilute} with $g_{11}>0$ and $g_{12}<0$ the mixed contributions are favored and $n_\text{dr}$ becomes much larger \cite{svistunov2015superfluid}: in that case vortices of different species can be tightly bound and act as stable dipoles with topological charge $(w_j^{(1)}=1,w_k^{(2)}=-1)$. 

The induced interaction is determined as the full interaction with the direct part canceled out,
\begin{multline}
p^{\text{eff}}_{11}\, e^{\,J_{11} C_{\vec{r}\vec{r}'}} = \\ \mathcal{Z}_\mathrm{top}^{-1}\Bigl[1  + \int_{L^2} \d^2 \vec s \int_{L^2} \d^2 \vec s'\, ( y_1^2 e^{ - J_{11}(C_{\vec{s}\vec{s}'}  - D_{\vec{rr'ss'}})} \\
+ y_2^2 e^{ -J_{22} C_{\vec{s}\vec{s}'} + J_{12}D_{\vec{rr'ss'}}} + \mathcal O(y_{1,2}^4)\Bigr].
\label{eq:p_pair} 
\end{multline}
Also the partition sum $\mathcal{Z}_\mathrm{top}$ is expanded to order $\mathcal O(y_{1,2}^2)$, where it consists of terms $e^{-J_{ii} C_{\vec{s}\vec{s}'}}$ for component $i$. Up to this order, the right-hand side of \eqref{eq:p_pair} can thus be written as
\begin{equation}
1 + y_1^2 \int \d^2 \vec{s'} \int \d^2 \vec{s} 
\left ( e^ { J_{11}\, D_{ \vec{rr'ss'}} } - 1\right )e^ {- J_{11} C_{\vec{ss'}}},
\label{eq:p_scr_2}
\end{equation}
plus an analogous contribution for $y_2^2$.

It the low-temperature limit, the most significant contributions to the partition function are those with tightly bound dipoles with small separation $\vec x\equiv\vec{s - s'}$, and we can use the dipole approximation. 
In terms of center-of-mass coordinates $(\vec{X},\vec{x})$ with $\vec X \equiv (\vec{s + s'})/2$, one can expand the dipole moment $D_\vec{rr'ss'}$ as
\begin{equation}
   D_{\vec{rr'ss'}} = \vec{x}\cdot \nabla \left(C_{\vec{rX}} - C_{\vec{r'X}}\right) + \mathcal{O}(x^3)
   \label{eq:dipole}
\end{equation}
where $\nabla \equiv \nabla_{\vec X}$; in the dipole approximation we retain only the linear term in $x$. The term in parenthesis in Eq.~\eqref{eq:p_scr_2} can thus be written as
\begin{multline}
    e^ {J_{11} D_{ \vec{rr'ss'}} } - 1 = \\
    J_{11} \,\vec{x}\cdot \nabla (C_{\vec{rX}} - C_{\vec{r'X}}) + \frac{1}{2}J_{11}^2 \left[\vec{x}\cdot \nabla (C_{\vec{rX}} - C_{\vec{r'X}}) \right]^2 + \mathcal O(x^3),
    \label{eq:D_taylor}
\end{multline}
and Eq.~\eqref{eq:p_scr_2} becomes
\begin{multline}
    1 + y_1^2 \int \d^2 \vec{X} \int \d^2 \vec{x} \, e^ {- J_{11}  C_{\vec{x}}}
    \Bigl( J_{11} \,\vec{x}\cdot \nabla (C_{\vec{rX}} - C_{\vec{r'X}})\\
    + \frac{1}{2}J_{11}^2 \left[\vec{x}\cdot \nabla (C_{\vec{rX}} - C_{\vec{r'X}}) \right]^2 \Bigr).
    \label{eq:p_scr_3}
\end{multline}
Upon angular integration over $\vec x$, the first term in \eqref{eq:p_scr_3} linear in $J_{11}$ vanishes since $\exp(-J_{11} C_\vec{x})$ does not depend on angle, and the second term yields
\begin{equation}
\int_0^{2\pi}\frac{\d\theta}{2\pi} \left[ \vec{x}\cdot \nabla (C_{\vec{rX}} - C_{\vec{r'X}}) \right]^2
= \frac{x^2}2 \abs{\nabla (C_{\vec{rX}} - C_{\vec{r'X}})}^2;
\label{eq:squareterm}
\end{equation}
the gradient term can be integrated by parts to give
\begin{align*}
   \int & \d^2 \vec{X}\, \abs{\nabla (C_{\vec{rX}} - C_{\vec{r'X}})}^2   \\ 
   &= - \int \d^2 \vec{X} \, \underbrace{ \nabla^2  [C_{\vec{rX}} - C_{\vec{r'X}}]}_{\delta(\vec{r - X}) - \delta(\vec{r' - X})}   \left (C_{\vec{rX}} - C_{\vec{r'X}}\right  ) \\
&= 2(C_{\vec{rr'}}- C_{\vec{0}}).
\end{align*}
The right-hand side of \eqref{eq:p_pair} then becomes, including the $y_2^2$ terms,
\begin{multline}
   1 + \pi (C_{\vec{rr'}}-C_\vec{0}) \\
   \times \int_0^\infty \d x \, x^3 \left( J_{11}^2 y_1^2 e^{-J_{11} C_x} + J_{12}^2 y_2^2 e^{- J_{22} C_x}  \right).
   \label{eq:before_reg}
\end{multline} 
The divergent contribution $C_{\vec 0}$ is regularized by setting a short-distance cutoff scale $a$ both for $\vec{r-r'}$ and $x$, and working with dimensionless lengths $\bar{\vec x}=\vec x/a$,
\begin{equation}
    C_\vec{x} - C_a = \ln(\abs{\vec{x}}/a)/(2\pi) = C_{\bar{\vec{x}}}.
    \label{eq:reg_a}
\end{equation}
After rescaling the integral and dropping the bars, we find the full interaction between two component-1 test charges given by
\begin{multline}
p^{\text{eff}}_{11}= e^{-J_{11} C_{\vec{rr'}}}\bigg( 1 + \pi C_{\vec{rr'}} \int_1^\infty \d x \, \big( J^2_{11} y_1^2 x^{3 - J_{11}/2\pi} \\ 
+ J_{12}^2 y_2^2 x^{3-J_{22}/2\pi}  \big) \bigg).
\end{multline}
Analogously, two component-2 test charges yield the corresponding interaction term
\begin{multline}
p^{\text{eff}}_{22} = e^{-J_{22} C_{\vec{rr'}}}\bigg( 1 + \pi C_{\vec{rr'}} \int_1^\infty \d x \, \big( J^2_{12} y_1^2 x^{3 - J_{11}/2\pi} \\
+ J_{22}^2 y_2^2 x^{3-J_{22}/2\pi}  \big) \bigg).
\end{multline}
Finally, the mixed case diagram in Fig.~\ref{fig:diagrams_y1}(c) has one test charge $\vec r$ of component 1, and the other $\vec r'$ of component 2. The interaction terms change slightly. Here, the only thing we have to change is that in \eqref{eq:dipole} there are two contributions with two different couplings,
\begin{multline}
    e^{J_{11}(C_{\vec{r s}} - C_{\vec{rs'}}) - J_{12}(C_{\vec{r's}} - C_{\vec{r's'}})} - 1 \\
    \shoveleft = \vec{x}\cdot \nabla ( J_{11}\, C_{\vec{rX}} - J_{12} \, C_{\vec{r'X}} ) \\
    + \frac12 \left[ \vec{x}\cdot \nabla ( J_{11}\, C_{\vec{rX}} - J_{12} \, C_{\vec{r'X}} ) \right]^2
    + \mathcal{O}(x^3).
\end{multline}
The subsequent steps proceed as above, and we find an expression similar to Eq.~\eqref{eq:before_reg} but with a factor $\pi(J_{11}J_{12}C_\vec{rr'} - (J_{11}^2+J_{12}^2)C_\vec{0}/2)$ before the integral. We can choose the same short-distance cutoff and thus obtain
\begin{multline}
p^{\text{eff}}_{12}= e^{- J_{12}C_{\vec{rr'}}}\bigg( 1 + \pi J_{12}C_{\vec{rr'}} 
 \int_1^\infty \d x \, \big( J_{11} y_1^2 x^{3 - J_{11}/2\pi} \\ + J_{22} y_2^2 x^{3-J_{22}/2\pi}  \big) \bigg).
\end{multline}
In all cases, the bare interaction term $e^{-J_{jk} C_{\vec{rr'}}}$ is screened by thermal fluctuations, and we can write the screened interaction as an effective direct interaction with renormalized coupling
\begin{align}
    J^\text{eff}_{11} & = J_{11} - \pi \int_1^\infty \d x \big( y_1^2 J_{11}^2 x^{3-J_{11}/2\pi} \notag \\
    & \qquad \qquad + y_2^2 J_{12}^2 x^{3-J_{22}/2\pi}\big), \\
    J^\text{eff}_{22} & = J_{22} - \pi \int_1^\infty \d x \big( y_1^2 J_{12}^2 x^{3-J_{11}/2\pi} \notag \\
    & \qquad \qquad + y_2^2 J_{22}^2 x^{3-J_{22}/2\pi}\big), \\
    J^\text{eff}_{12} & = J_{12} - \pi J_{12}\int_1^\infty \d x\,\big( y_{1}^2 
    J_{11} x^{3-J_{11}/2\pi} \notag \\
    & \qquad \qquad + y_{2}^2 J_{22} x^{3-J_{22}/2\pi}\big).
    \label{eq:effective_couplings}
\end{align}
Since the couplings depend on each other, we can solve this set of equations using a flow equation for the three effective couplings as a function of scale. This is done by splitting the integrals $\int_1^\infty = \int_1^b + \int_b^\infty$ and introducing the new intermediate couplings $\tilde J$ which include the fluctuations in the range $x=1\dotsc b$, such that to order $y^2$ one finds
\begin{align}
    J^{-1}_\text{eff} &= \tilde{J}^{-1} +  \pi \, y^2 \int_b^{\infty} \d x \, x^{3 - J/2\pi} + \mathcal{O}(y^4) \label{eq:renorm} \\
    \tilde{J}^{-1} &= J^{-1} + \pi \, y^2 \int_1^b \d x \, x^{3- J/2\pi}.
\end{align}
If we now express $y$ in terms of the rescaled $\tilde y= b^{2 - J/4\pi} y$, the integration variable in \eqref{eq:renorm} can be rescaled back to the original range $x=1\dotsc\infty$ and we obtain the same form \eqref{eq:effective_couplings} as before, but with rescaled couplings. An infinitesimal rescaling $b=e^l\approx1+l$ for $l\ll1$ immediately yields the five coupled renormalization group equations \eqref{eq:RG}.

\bibliography{./Bibliography.bib}

\begin{thebibliography}{71}%
\makeatletter
\providecommand \@ifxundefined [1]{%
 \@ifx{#1\undefined}
}%
\providecommand \@ifnum [1]{%
 \ifnum #1\expandafter \@firstoftwo
 \else \expandafter \@secondoftwo
 \fi
}%
\providecommand \@ifx [1]{%
 \ifx #1\expandafter \@firstoftwo
 \else \expandafter \@secondoftwo
 \fi
}%
\providecommand \natexlab [1]{#1}%
\providecommand \enquote  [1]{``#1''}%
\providecommand \bibnamefont  [1]{#1}%
\providecommand \bibfnamefont [1]{#1}%
\providecommand \citenamefont [1]{#1}%
\providecommand \href@noop [0]{\@secondoftwo}%
\providecommand \href [0]{\begingroup \@sanitize@url \@href}%
\providecommand \@href[1]{\@@startlink{#1}\@@href}%
\providecommand \@@href[1]{\endgroup#1\@@endlink}%
\providecommand \@sanitize@url [0]{\catcode `\\12\catcode `\$12\catcode
  `\&12\catcode `\#12\catcode `\^12\catcode `\_12\catcode `\%12\relax}%
\providecommand \@@startlink[1]{}%
\providecommand \@@endlink[0]{}%
\providecommand \url  [0]{\begingroup\@sanitize@url \@url }%
\providecommand \@url [1]{\endgroup\@href {#1}{\urlprefix }}%
\providecommand \urlprefix  [0]{URL }%
\providecommand \Eprint [0]{\href }%
\providecommand \doibase [0]{http://dx.doi.org/}%
\providecommand \selectlanguage [0]{\@gobble}%
\providecommand \bibinfo  [0]{\@secondoftwo}%
\providecommand \bibfield  [0]{\@secondoftwo}%
\providecommand \translation [1]{[#1]}%
\providecommand \BibitemOpen [0]{}%
\providecommand \bibitemStop [0]{}%
\providecommand \bibitemNoStop [0]{.\EOS\space}%
\providecommand \EOS [0]{\spacefactor3000\relax}%
\providecommand \BibitemShut  [1]{\csname bibitem#1\endcsname}%
\let\auto@bib@innerbib\@empty
\bibitem [{\citenamefont {Larsen}(1963)}]{larsen1963binary}%
  \BibitemOpen
  \bibfield  {author} {\bibinfo {author} {\bibfnamefont {D.~M.}\ \bibnamefont
  {Larsen}},\ }\href {\doibase 10.1016/0003-4916(63)90066-6} {\bibfield
  {journal} {\bibinfo  {journal} {Ann. Phys. (N.Y.)}\ }\textbf {\bibinfo
  {volume} {24}},\ \bibinfo {pages} {89} (\bibinfo {year} {1963})}\BibitemShut
  {NoStop}%
\bibitem [{\citenamefont {Graf}\ \emph {et~al.}(1967)\citenamefont {Graf},
  \citenamefont {Lee},\ and\ \citenamefont {Reppy}}]{graf1967phase}%
  \BibitemOpen
  \bibfield  {author} {\bibinfo {author} {\bibfnamefont {E.~H.}\ \bibnamefont
  {Graf}}, \bibinfo {author} {\bibfnamefont {D.~M.}\ \bibnamefont {Lee}}, \
  and\ \bibinfo {author} {\bibfnamefont {J.~D.}\ \bibnamefont {Reppy}},\
  }\href@noop {} {\bibfield  {journal} {\bibinfo  {journal} {Phys. Rev. Lett.}\
  }\textbf {\bibinfo {volume} {19}},\ \bibinfo {pages} {417} (\bibinfo {year}
  {1967})}\BibitemShut {NoStop}%
\bibitem [{\citenamefont {Andreev}\ and\ \citenamefont
  {Bashkin}(1975)}]{andreev1975af}%
  \BibitemOpen
  \bibfield  {author} {\bibinfo {author} {\bibfnamefont {A.~F.}\ \bibnamefont
  {Andreev}}\ and\ \bibinfo {author} {\bibfnamefont {E.~P.}\ \bibnamefont
  {Bashkin}},\ }\href@noop {} {\bibfield  {journal} {\bibinfo  {journal} {Zh.
  Eksp. Teor. Fiz.}\ }\textbf {\bibinfo {volume} {69}},\ \bibinfo {pages} {319}
  (\bibinfo {year} {1975})},\ \bibinfo {note} {[Sov. Phys. JETP 42, 164
  (1975)]}\BibitemShut {NoStop}%
\bibitem [{\citenamefont {Ho}\ and\ \citenamefont
  {Shenoy}(1996)}]{ho1996binary}%
  \BibitemOpen
  \bibfield  {author} {\bibinfo {author} {\bibfnamefont {T.-L.}\ \bibnamefont
  {Ho}}\ and\ \bibinfo {author} {\bibfnamefont {V.~B.}\ \bibnamefont
  {Shenoy}},\ }\href@noop {} {\bibfield  {journal} {\bibinfo  {journal} {Phys.
  Rev. Lett.}\ }\textbf {\bibinfo {volume} {77}},\ \bibinfo {pages} {3276}
  (\bibinfo {year} {1996})}\BibitemShut {NoStop}%
\bibitem [{\citenamefont {Myatt}\ \emph {et~al.}(1997)\citenamefont {Myatt},
  \citenamefont {Burt}, \citenamefont {Ghrist}, \citenamefont {Cornell},\ and\
  \citenamefont {Wieman}}]{myatt1997production}%
  \BibitemOpen
  \bibfield  {author} {\bibinfo {author} {\bibfnamefont {C.~J.}\ \bibnamefont
  {Myatt}}, \bibinfo {author} {\bibfnamefont {E.~A.}\ \bibnamefont {Burt}},
  \bibinfo {author} {\bibfnamefont {R.~W.}\ \bibnamefont {Ghrist}}, \bibinfo
  {author} {\bibfnamefont {E.~A.}\ \bibnamefont {Cornell}}, \ and\ \bibinfo
  {author} {\bibfnamefont {C.~E.}\ \bibnamefont {Wieman}},\ }\href@noop {}
  {\bibfield  {journal} {\bibinfo  {journal} {Phys. Rev. Lett.}\ }\textbf
  {\bibinfo {volume} {78}},\ \bibinfo {pages} {586} (\bibinfo {year}
  {1997})}\BibitemShut {NoStop}%
\bibitem [{\citenamefont {Altman}\ \emph {et~al.}(2003)\citenamefont {Altman},
  \citenamefont {Hofstetter}, \citenamefont {Demler},\ and\ \citenamefont
  {Lukin}}]{altman2003phase}%
  \BibitemOpen
  \bibfield  {author} {\bibinfo {author} {\bibfnamefont {E.}~\bibnamefont
  {Altman}}, \bibinfo {author} {\bibfnamefont {W.}~\bibnamefont {Hofstetter}},
  \bibinfo {author} {\bibfnamefont {E.}~\bibnamefont {Demler}}, \ and\ \bibinfo
  {author} {\bibfnamefont {M.~D.}\ \bibnamefont {Lukin}},\ }\href@noop {}
  {\bibfield  {journal} {\bibinfo  {journal} {New J. Phys.}\ }\textbf {\bibinfo
  {volume} {5}},\ \bibinfo {pages} {113} (\bibinfo {year} {2003})}\BibitemShut
  {NoStop}%
\bibitem [{\citenamefont {Stamper-Kurn}\ and\ \citenamefont
  {Ueda}(2013)}]{stamper2013spinor}%
  \BibitemOpen
  \bibfield  {author} {\bibinfo {author} {\bibfnamefont {D.~M.}\ \bibnamefont
  {Stamper-Kurn}}\ and\ \bibinfo {author} {\bibfnamefont {M.}~\bibnamefont
  {Ueda}},\ }\href@noop {} {\bibfield  {journal} {\bibinfo  {journal} {Rev.
  Mod. Phys.}\ }\textbf {\bibinfo {volume} {85}},\ \bibinfo {pages} {1191}
  (\bibinfo {year} {2013})}\BibitemShut {NoStop}%
\bibitem [{\citenamefont {Fil}\ and\ \citenamefont
  {Shevchenko}(2004)}]{fil2004drag}%
  \BibitemOpen
  \bibfield  {author} {\bibinfo {author} {\bibfnamefont {D.~V.}\ \bibnamefont
  {Fil}}\ and\ \bibinfo {author} {\bibfnamefont {S.~I.}\ \bibnamefont
  {Shevchenko}},\ }\href@noop {} {\bibfield  {journal} {\bibinfo  {journal}
  {Low Temp. Phys.}\ }\textbf {\bibinfo {volume} {30}},\ \bibinfo {pages} {770}
  (\bibinfo {year} {2004})}\BibitemShut {NoStop}%
\bibitem [{\citenamefont {Szab{\'o}}\ \emph {et~al.}(2001)\citenamefont
  {Szab{\'o}}, \citenamefont {Samuely}, \citenamefont
  {Ka{\v{c}}mar{\v{c}}{\'\i}k}, \citenamefont {Klein}, \citenamefont {Marcus},
  \citenamefont {Fruchart}, \citenamefont {Miraglia}, \citenamefont
  {Marcenat},\ and\ \citenamefont {Jansen}}]{szabo2001evidence}%
  \BibitemOpen
  \bibfield  {author} {\bibinfo {author} {\bibfnamefont {P.}~\bibnamefont
  {Szab{\'o}}}, \bibinfo {author} {\bibfnamefont {P.}~\bibnamefont {Samuely}},
  \bibinfo {author} {\bibfnamefont {J.}~\bibnamefont
  {Ka{\v{c}}mar{\v{c}}{\'\i}k}}, \bibinfo {author} {\bibfnamefont
  {T.}~\bibnamefont {Klein}}, \bibinfo {author} {\bibfnamefont
  {J.}~\bibnamefont {Marcus}}, \bibinfo {author} {\bibfnamefont
  {D.}~\bibnamefont {Fruchart}}, \bibinfo {author} {\bibfnamefont
  {S.}~\bibnamefont {Miraglia}}, \bibinfo {author} {\bibfnamefont
  {C.}~\bibnamefont {Marcenat}}, \ and\ \bibinfo {author} {\bibfnamefont
  {A.~G.~M.}\ \bibnamefont {Jansen}},\ }\href {\doibase
  10.1103/PhysRevLett.87.137005} {\bibfield  {journal} {\bibinfo  {journal}
  {Phys. Rev. Lett.}\ }\textbf {\bibinfo {volume} {87}},\ \bibinfo {pages}
  {137005} (\bibinfo {year} {2001})}\BibitemShut {NoStop}%
\bibitem [{\citenamefont {Petrov}(2015)}]{petrov2015quantum}%
  \BibitemOpen
  \bibfield  {author} {\bibinfo {author} {\bibfnamefont {D.~S.}\ \bibnamefont
  {Petrov}},\ }\href@noop {} {\bibfield  {journal} {\bibinfo  {journal} {Phys.
  Rev. Lett.}\ }\textbf {\bibinfo {volume} {115}},\ \bibinfo {pages} {155302}
  (\bibinfo {year} {2015})}\BibitemShut {NoStop}%
\bibitem [{\citenamefont {Cheiney}\ \emph {et~al.}(2018)\citenamefont
  {Cheiney}, \citenamefont {Cabrera}, \citenamefont {Sanz}, \citenamefont
  {Naylor}, \citenamefont {Tanzi},\ and\ \citenamefont
  {Tarruell}}]{cheiney2018bright}%
  \BibitemOpen
  \bibfield  {author} {\bibinfo {author} {\bibfnamefont {P.}~\bibnamefont
  {Cheiney}}, \bibinfo {author} {\bibfnamefont {C.~R.}\ \bibnamefont
  {Cabrera}}, \bibinfo {author} {\bibfnamefont {J.}~\bibnamefont {Sanz}},
  \bibinfo {author} {\bibfnamefont {B.}~\bibnamefont {Naylor}}, \bibinfo
  {author} {\bibfnamefont {L.}~\bibnamefont {Tanzi}}, \ and\ \bibinfo {author}
  {\bibfnamefont {L.}~\bibnamefont {Tarruell}},\ }\href@noop {} {\bibfield
  {journal} {\bibinfo  {journal} {Phys. Rev. Lett.}\ }\textbf {\bibinfo
  {volume} {120}},\ \bibinfo {pages} {135301} (\bibinfo {year}
  {2018})}\BibitemShut {NoStop}%
\bibitem [{\citenamefont {Cabrera}\ \emph {et~al.}(2018)\citenamefont
  {Cabrera}, \citenamefont {Tanzi}, \citenamefont {Sanz}, \citenamefont
  {Naylor}, \citenamefont {Thomas}, \citenamefont {Cheiney},\ and\
  \citenamefont {Tarruell}}]{cabrera2018quantum}%
  \BibitemOpen
  \bibfield  {author} {\bibinfo {author} {\bibfnamefont {C.~R.}\ \bibnamefont
  {Cabrera}}, \bibinfo {author} {\bibfnamefont {L.}~\bibnamefont {Tanzi}},
  \bibinfo {author} {\bibfnamefont {J.}~\bibnamefont {Sanz}}, \bibinfo {author}
  {\bibfnamefont {B.}~\bibnamefont {Naylor}}, \bibinfo {author} {\bibfnamefont
  {P.}~\bibnamefont {Thomas}}, \bibinfo {author} {\bibfnamefont
  {P.}~\bibnamefont {Cheiney}}, \ and\ \bibinfo {author} {\bibfnamefont
  {L.}~\bibnamefont {Tarruell}},\ }\href {\doibase 10.1126/science.aao5686}
  {\bibfield  {journal} {\bibinfo  {journal} {Science}\ }\textbf {\bibinfo
  {volume} {359}},\ \bibinfo {pages} {301} (\bibinfo {year}
  {2018})}\BibitemShut {NoStop}%
\bibitem [{\citenamefont {Ye}\ \emph {et~al.}(2018)\citenamefont {Ye},
  \citenamefont {Huang}, \citenamefont {Zhuang}, \citenamefont {Zhong},\ and\
  \citenamefont {Lee}}]{ye2018dressed}%
  \BibitemOpen
  \bibfield  {author} {\bibinfo {author} {\bibfnamefont {Q.}~\bibnamefont
  {Ye}}, \bibinfo {author} {\bibfnamefont {J.}~\bibnamefont {Huang}}, \bibinfo
  {author} {\bibfnamefont {M.}~\bibnamefont {Zhuang}}, \bibinfo {author}
  {\bibfnamefont {H.}~\bibnamefont {Zhong}}, \ and\ \bibinfo {author}
  {\bibfnamefont {C.}~\bibnamefont {Lee}},\ }\href@noop {} {\bibfield
  {journal} {\bibinfo  {journal} {Sci. Rep.}\ }\textbf {\bibinfo {volume}
  {8}},\ \bibinfo {pages} {4484} (\bibinfo {year} {2018})}\BibitemShut
  {NoStop}%
\bibitem [{\citenamefont {Schulze}\ \emph {et~al.}(2018)\citenamefont
  {Schulze}, \citenamefont {Hartmann}, \citenamefont {Voges}, \citenamefont
  {Gempel}, \citenamefont {Tiemann}, \citenamefont {Zenesini},\ and\
  \citenamefont {Ospelkaus}}]{schulze2018feshbach}%
  \BibitemOpen
  \bibfield  {author} {\bibinfo {author} {\bibfnamefont {T.~A.}\ \bibnamefont
  {Schulze}}, \bibinfo {author} {\bibfnamefont {T.}~\bibnamefont {Hartmann}},
  \bibinfo {author} {\bibfnamefont {K.~K.}\ \bibnamefont {Voges}}, \bibinfo
  {author} {\bibfnamefont {M.~W.}\ \bibnamefont {Gempel}}, \bibinfo {author}
  {\bibfnamefont {E.}~\bibnamefont {Tiemann}}, \bibinfo {author} {\bibfnamefont
  {A.}~\bibnamefont {Zenesini}}, \ and\ \bibinfo {author} {\bibfnamefont
  {S.}~\bibnamefont {Ospelkaus}},\ }\href {\doibase 10.1103/PhysRevA.97.023623}
  {\bibfield  {journal} {\bibinfo  {journal} {Phys. Rev. A}\ }\textbf {\bibinfo
  {volume} {97}},\ \bibinfo {pages} {023623} (\bibinfo {year}
  {2018})}\BibitemShut {NoStop}%
\bibitem [{\citenamefont {Petrov}\ \emph
  {et~al.}(2000{\natexlab{a}})\citenamefont {Petrov}, \citenamefont
  {Shlyapnikov},\ and\ \citenamefont {Walraven}}]{petrov2000regimes}%
  \BibitemOpen
  \bibfield  {author} {\bibinfo {author} {\bibfnamefont {D.~S.}\ \bibnamefont
  {Petrov}}, \bibinfo {author} {\bibfnamefont {G.~V.}\ \bibnamefont
  {Shlyapnikov}}, \ and\ \bibinfo {author} {\bibfnamefont {J.~T.~M.}\
  \bibnamefont {Walraven}},\ }\href@noop {} {\bibfield  {journal} {\bibinfo
  {journal} {Phys. Rev. Lett.}\ }\textbf {\bibinfo {volume} {85}},\ \bibinfo
  {pages} {3745} (\bibinfo {year} {2000}{\natexlab{a}})}\BibitemShut {NoStop}%
\bibitem [{\citenamefont {Petrov}\ and\ \citenamefont
  {Astrakharchik}(2016)}]{petrov2016ultradilute}%
  \BibitemOpen
  \bibfield  {author} {\bibinfo {author} {\bibfnamefont {D.~S.}\ \bibnamefont
  {Petrov}}\ and\ \bibinfo {author} {\bibfnamefont {G.~E.}\ \bibnamefont
  {Astrakharchik}},\ }\href {\doibase 10.1103/PhysRevLett.117.100401}
  {\bibfield  {journal} {\bibinfo  {journal} {Phys. Rev. Lett.}\ }\textbf
  {\bibinfo {volume} {117}},\ \bibinfo {pages} {100401} (\bibinfo {year}
  {2016})}\BibitemShut {NoStop}%
\bibitem [{\citenamefont {Busch}\ \emph {et~al.}(1997)\citenamefont {Busch},
  \citenamefont {Cirac}, \citenamefont {Perez-Garcia},\ and\ \citenamefont
  {Zoller}}]{busch1997stability}%
  \BibitemOpen
  \bibfield  {author} {\bibinfo {author} {\bibfnamefont {T.}~\bibnamefont
  {Busch}}, \bibinfo {author} {\bibfnamefont {J.~I.}\ \bibnamefont {Cirac}},
  \bibinfo {author} {\bibfnamefont {V.~M.}\ \bibnamefont {Perez-Garcia}}, \
  and\ \bibinfo {author} {\bibfnamefont {P.}~\bibnamefont {Zoller}},\
  }\href@noop {} {\bibfield  {journal} {\bibinfo  {journal} {Phys. Rev. A}\
  }\textbf {\bibinfo {volume} {56}},\ \bibinfo {pages} {2978} (\bibinfo {year}
  {1997})}\BibitemShut {NoStop}%
\bibitem [{\citenamefont {Bashkin}\ and\ \citenamefont
  {Vagov}(1997)}]{bashkin1997instability}%
  \BibitemOpen
  \bibfield  {author} {\bibinfo {author} {\bibfnamefont {E.~P.}\ \bibnamefont
  {Bashkin}}\ and\ \bibinfo {author} {\bibfnamefont {A.~V.}\ \bibnamefont
  {Vagov}},\ }\href@noop {} {\bibfield  {journal} {\bibinfo  {journal} {Phys.
  Rev. B}\ }\textbf {\bibinfo {volume} {56}},\ \bibinfo {pages} {6207}
  (\bibinfo {year} {1997})}\BibitemShut {NoStop}%
\bibitem [{\citenamefont {Ao}\ and\ \citenamefont {Chui}(2000)}]{ao2000two}%
  \BibitemOpen
  \bibfield  {author} {\bibinfo {author} {\bibfnamefont {P.}~\bibnamefont
  {Ao}}\ and\ \bibinfo {author} {\bibfnamefont {S.~T.}\ \bibnamefont {Chui}},\
  }\href@noop {} {\bibfield  {journal} {\bibinfo  {journal} {J. Phys. B}\
  }\textbf {\bibinfo {volume} {33}},\ \bibinfo {pages} {535} (\bibinfo {year}
  {2000})}\BibitemShut {NoStop}%
\bibitem [{\citenamefont {Pethick}\ and\ \citenamefont
  {Smith}(2008)}]{pethick2002bose}%
  \BibitemOpen
  \bibfield  {author} {\bibinfo {author} {\bibfnamefont {C.~J.}\ \bibnamefont
  {Pethick}}\ and\ \bibinfo {author} {\bibfnamefont {H.}~\bibnamefont
  {Smith}},\ }\href@noop {} {\emph {\bibinfo {title} {Bose-Einstein
  Condensation in Dilute Gases}}},\ \bibinfo {edition} {2nd}\ ed.\ (\bibinfo
  {publisher} {Cambridge University Press},\ \bibinfo {year}
  {2008})\BibitemShut {NoStop}%
\bibitem [{\citenamefont {Pitaevskii}\ and\ \citenamefont
  {Stringari}(2016)}]{pitaevskii2016bose}%
  \BibitemOpen
  \bibfield  {author} {\bibinfo {author} {\bibfnamefont {L.}~\bibnamefont
  {Pitaevskii}}\ and\ \bibinfo {author} {\bibfnamefont {S.}~\bibnamefont
  {Stringari}},\ }\href@noop {} {\emph {\bibinfo {title} {Bose-Einstein
  Condensation and Superfluidity}}}\ (\bibinfo  {publisher} {Oxford University
  Press},\ \bibinfo {year} {2016})\BibitemShut {NoStop}%
\bibitem [{\citenamefont {Fil}\ and\ \citenamefont
  {Shevchenko}(2005)}]{fil2005nondissipative}%
  \BibitemOpen
  \bibfield  {author} {\bibinfo {author} {\bibfnamefont {D.~V.}\ \bibnamefont
  {Fil}}\ and\ \bibinfo {author} {\bibfnamefont {S.~I.}\ \bibnamefont
  {Shevchenko}},\ }\href@noop {} {\bibfield  {journal} {\bibinfo  {journal}
  {Phys. Rev. A}\ }\textbf {\bibinfo {volume} {72}},\ \bibinfo {pages} {013616}
  (\bibinfo {year} {2005})}\BibitemShut {NoStop}%
\bibitem [{\citenamefont {Ishino}\ \emph {et~al.}(2011)\citenamefont {Ishino},
  \citenamefont {Tsubota},\ and\ \citenamefont
  {Takeuchi}}]{ishino2011countersuperflow}%
  \BibitemOpen
  \bibfield  {author} {\bibinfo {author} {\bibfnamefont {S.}~\bibnamefont
  {Ishino}}, \bibinfo {author} {\bibfnamefont {M.}~\bibnamefont {Tsubota}}, \
  and\ \bibinfo {author} {\bibfnamefont {H.}~\bibnamefont {Takeuchi}},\ }\href
  {\doibase 10.1103/PhysRevA.83.063602} {\bibfield  {journal} {\bibinfo
  {journal} {Phys. Rev. A}\ }\textbf {\bibinfo {volume} {83}},\ \bibinfo
  {pages} {063602} (\bibinfo {year} {2011})}\BibitemShut {NoStop}%
\bibitem [{\citenamefont {Hofer}\ \emph {et~al.}(2012)\citenamefont {Hofer},
  \citenamefont {Bruder},\ and\ \citenamefont
  {Stojanovi{\'c}}}]{hofer2012superfluid}%
  \BibitemOpen
  \bibfield  {author} {\bibinfo {author} {\bibfnamefont {P.~P.}\ \bibnamefont
  {Hofer}}, \bibinfo {author} {\bibfnamefont {C.}~\bibnamefont {Bruder}}, \
  and\ \bibinfo {author} {\bibfnamefont {V.~M.}\ \bibnamefont
  {Stojanovi{\'c}}},\ }\href@noop {} {\bibfield  {journal} {\bibinfo  {journal}
  {Phys. Rev. A}\ }\textbf {\bibinfo {volume} {86}},\ \bibinfo {pages} {033627}
  (\bibinfo {year} {2012})}\BibitemShut {NoStop}%
\bibitem [{\citenamefont {Nespolo}\ \emph {et~al.}(2017)\citenamefont
  {Nespolo}, \citenamefont {Astrakharchik},\ and\ \citenamefont
  {Recati}}]{nespolo2017andreev}%
  \BibitemOpen
  \bibfield  {author} {\bibinfo {author} {\bibfnamefont {J.}~\bibnamefont
  {Nespolo}}, \bibinfo {author} {\bibfnamefont {G.~E.}\ \bibnamefont
  {Astrakharchik}}, \ and\ \bibinfo {author} {\bibfnamefont {A.}~\bibnamefont
  {Recati}},\ }\href@noop {} {\bibfield  {journal} {\bibinfo  {journal} {New J.
  Phys.}\ }\textbf {\bibinfo {volume} {19}},\ \bibinfo {pages} {125005}
  (\bibinfo {year} {2017})}\BibitemShut {NoStop}%
\bibitem [{\citenamefont {Svistunov}\ \emph {et~al.}(2015)\citenamefont
  {Svistunov}, \citenamefont {Babaev},\ and\ \citenamefont
  {Prokof'ev}}]{svistunov2015superfluid}%
  \BibitemOpen
  \bibfield  {author} {\bibinfo {author} {\bibfnamefont {B.~V.}\ \bibnamefont
  {Svistunov}}, \bibinfo {author} {\bibfnamefont {E.~S.}\ \bibnamefont
  {Babaev}}, \ and\ \bibinfo {author} {\bibfnamefont {N.~V.}\ \bibnamefont
  {Prokof'ev}},\ }\href@noop {} {\emph {\bibinfo {title} {Superfluid States of
  Matter}}}\ (\bibinfo  {publisher} {CRC Press},\ \bibinfo {year}
  {2015})\BibitemShut {NoStop}%
\bibitem [{\citenamefont {Parisi}\ \emph {et~al.}(2018)\citenamefont {Parisi},
  \citenamefont {Astrakharchik},\ and\ \citenamefont
  {Giorgini}}]{parisi2018spin}%
  \BibitemOpen
  \bibfield  {author} {\bibinfo {author} {\bibfnamefont {L.}~\bibnamefont
  {Parisi}}, \bibinfo {author} {\bibfnamefont {G.~E.}\ \bibnamefont
  {Astrakharchik}}, \ and\ \bibinfo {author} {\bibfnamefont {S.}~\bibnamefont
  {Giorgini}},\ }\href@noop {} {\bibfield  {journal} {\bibinfo  {journal}
  {Phys. Rev. Lett.}\ }\textbf {\bibinfo {volume} {121}},\ \bibinfo {pages}
  {025302} (\bibinfo {year} {2018})}\BibitemShut {NoStop}%
\bibitem [{\citenamefont {Sellin}\ and\ \citenamefont
  {Babaev}(2018)}]{sellin2018}%
  \BibitemOpen
  \bibfield  {author} {\bibinfo {author} {\bibfnamefont {K.}~\bibnamefont
  {Sellin}}\ and\ \bibinfo {author} {\bibfnamefont {E.}~\bibnamefont
  {Babaev}},\ }\href@noop {} {\bibfield  {journal} {\bibinfo  {journal} {Phys.
  Rev. B}\ }\textbf {\bibinfo {volume} {97}},\ \bibinfo {pages} {094517}
  (\bibinfo {year} {2018})}\BibitemShut {NoStop}%
\bibitem [{\citenamefont {Konietin}\ and\ \citenamefont
  {Pastukhov}(2018)}]{Konietin2018}%
  \BibitemOpen
  \bibfield  {author} {\bibinfo {author} {\bibfnamefont {P.}~\bibnamefont
  {Konietin}}\ and\ \bibinfo {author} {\bibfnamefont {V.}~\bibnamefont
  {Pastukhov}},\ }\href@noop {} {\bibfield  {journal} {\bibinfo  {journal} {J.
  Low Temp. Phys.}\ }\textbf {\bibinfo {volume} {190}},\ \bibinfo {pages} {256}
  (\bibinfo {year} {2018})}\BibitemShut {NoStop}%
\bibitem [{\citenamefont {Berezinskii}(1972)}]{berezinskii1972destruction}%
  \BibitemOpen
  \bibfield  {author} {\bibinfo {author} {\bibfnamefont {V.~L.}\ \bibnamefont
  {Berezinskii}},\ }\href@noop {} {\bibfield  {journal} {\bibinfo  {journal}
  {Zh. Eksp. Teor. Fiz.}\ }\textbf {\bibinfo {volume} {61}},\ \bibinfo {pages}
  {1144} (\bibinfo {year} {1972})},\ \bibinfo {note} {[Sov. Phys. JETP 34, 610
  (1971)]}\BibitemShut {NoStop}%
\bibitem [{\citenamefont {Kosterlitz}\ and\ \citenamefont
  {Thouless}(1973)}]{kosterlitz1973ordering}%
  \BibitemOpen
  \bibfield  {author} {\bibinfo {author} {\bibfnamefont {J.~M.}\ \bibnamefont
  {Kosterlitz}}\ and\ \bibinfo {author} {\bibfnamefont {D.~J.}\ \bibnamefont
  {Thouless}},\ }\href@noop {} {\bibfield  {journal} {\bibinfo  {journal} {J.
  Phys. C: Solid State Phys.}\ }\textbf {\bibinfo {volume} {6}},\ \bibinfo
  {pages} {1181} (\bibinfo {year} {1973})}\BibitemShut {NoStop}%
\bibitem [{\citenamefont {Kosterlitz}(1974)}]{kosterlitz1974critical}%
  \BibitemOpen
  \bibfield  {author} {\bibinfo {author} {\bibfnamefont {J.~M.}\ \bibnamefont
  {Kosterlitz}},\ }\href@noop {} {\bibfield  {journal} {\bibinfo  {journal} {J.
  Phys. C: Solid State Phys.}\ }\textbf {\bibinfo {volume} {7}},\ \bibinfo
  {pages} {1046} (\bibinfo {year} {1974})}\BibitemShut {NoStop}%
\bibitem [{\citenamefont {Dahl}\ \emph {et~al.}(2008)\citenamefont {Dahl},
  \citenamefont {Babaev}, \citenamefont {Kragset},\ and\ \citenamefont
  {Sudb{\o}}}]{dahl2008}%
  \BibitemOpen
  \bibfield  {author} {\bibinfo {author} {\bibfnamefont {E.~K.}\ \bibnamefont
  {Dahl}}, \bibinfo {author} {\bibfnamefont {E.}~\bibnamefont {Babaev}},
  \bibinfo {author} {\bibfnamefont {S.}~\bibnamefont {Kragset}}, \ and\
  \bibinfo {author} {\bibfnamefont {A.}~\bibnamefont {Sudb{\o}}},\ }\href@noop
  {} {\bibfield  {journal} {\bibinfo  {journal} {Phys. Rev. B}\ }\textbf
  {\bibinfo {volume} {77}},\ \bibinfo {pages} {144519} (\bibinfo {year}
  {2008})}\BibitemShut {NoStop}%
\bibitem [{\citenamefont {Hadzibabic}\ \emph {et~al.}(2006)\citenamefont
  {Hadzibabic}, \citenamefont {Kr{\"u}ger}, \citenamefont {Cheneau},
  \citenamefont {Battelier},\ and\ \citenamefont
  {Dalibard}}]{hadzibabic2006berezinskii}%
  \BibitemOpen
  \bibfield  {author} {\bibinfo {author} {\bibfnamefont {Z.}~\bibnamefont
  {Hadzibabic}}, \bibinfo {author} {\bibfnamefont {P.}~\bibnamefont
  {Kr{\"u}ger}}, \bibinfo {author} {\bibfnamefont {M.}~\bibnamefont {Cheneau}},
  \bibinfo {author} {\bibfnamefont {B.}~\bibnamefont {Battelier}}, \ and\
  \bibinfo {author} {\bibfnamefont {J.}~\bibnamefont {Dalibard}},\ }\href@noop
  {} {\bibfield  {journal} {\bibinfo  {journal} {Nature (London)}\ }\textbf
  {\bibinfo {volume} {441}},\ \bibinfo {pages} {1118} (\bibinfo {year}
  {2006})}\BibitemShut {NoStop}%
\bibitem [{\citenamefont {Ran{\c{c}}on}\ and\ \citenamefont
  {Dupuis}(2017)}]{rancon2017kosterlitz}%
  \BibitemOpen
  \bibfield  {author} {\bibinfo {author} {\bibfnamefont {A.}~\bibnamefont
  {Ran{\c{c}}on}}\ and\ \bibinfo {author} {\bibfnamefont {N.}~\bibnamefont
  {Dupuis}},\ }\href@noop {} {\bibfield  {journal} {\bibinfo  {journal} {Phys.
  Rev. B}\ }\textbf {\bibinfo {volume} {96}},\ \bibinfo {pages} {214512}
  (\bibinfo {year} {2017})}\BibitemShut {NoStop}%
\bibitem [{\citenamefont {Takeuchi}\ \emph {et~al.}(2010)\citenamefont
  {Takeuchi}, \citenamefont {Ishino},\ and\ \citenamefont
  {Tsubota}}]{takeuchi2010binary}%
  \BibitemOpen
  \bibfield  {author} {\bibinfo {author} {\bibfnamefont {H.}~\bibnamefont
  {Takeuchi}}, \bibinfo {author} {\bibfnamefont {S.}~\bibnamefont {Ishino}}, \
  and\ \bibinfo {author} {\bibfnamefont {M.}~\bibnamefont {Tsubota}},\ }\href
  {\doibase 10.1103/PhysRevLett.105.205301} {\bibfield  {journal} {\bibinfo
  {journal} {Phys. Rev. Lett.}\ }\textbf {\bibinfo {volume} {105}},\ \bibinfo
  {pages} {205301} (\bibinfo {year} {2010})}\BibitemShut {NoStop}%
\bibitem [{\citenamefont {Karl}\ \emph {et~al.}(2013)\citenamefont {Karl},
  \citenamefont {Nowak},\ and\ \citenamefont {Gasenzer}}]{karl2013universal}%
  \BibitemOpen
  \bibfield  {author} {\bibinfo {author} {\bibfnamefont {M.}~\bibnamefont
  {Karl}}, \bibinfo {author} {\bibfnamefont {B.}~\bibnamefont {Nowak}}, \ and\
  \bibinfo {author} {\bibfnamefont {T.}~\bibnamefont {Gasenzer}},\ }\href
  {\doibase 10.1103/PhysRevA.88.063615} {\bibfield  {journal} {\bibinfo
  {journal} {Phys. Rev. A}\ }\textbf {\bibinfo {volume} {88}},\ \bibinfo
  {pages} {063615} (\bibinfo {year} {2013})}\BibitemShut {NoStop}%
\bibitem [{\citenamefont {Karl}\ and\ \citenamefont
  {Gasenzer}(2017)}]{karl2017strongly}%
  \BibitemOpen
  \bibfield  {author} {\bibinfo {author} {\bibfnamefont {M.}~\bibnamefont
  {Karl}}\ and\ \bibinfo {author} {\bibfnamefont {T.}~\bibnamefont
  {Gasenzer}},\ }\href@noop {} {\bibfield  {journal} {\bibinfo  {journal} {New
  J. Phys.}\ }\textbf {\bibinfo {volume} {19}},\ \bibinfo {pages} {093014}
  (\bibinfo {year} {2017})}\BibitemShut {NoStop}%
\bibitem [{\citenamefont {Gallem{\'\i}}\ \emph {et~al.}(2018)\citenamefont
  {Gallem{\'\i}}, \citenamefont {Pitaevskii}, \citenamefont {Stringari},\ and\
  \citenamefont {Recati}}]{gallemi2018magnetic}%
  \BibitemOpen
  \bibfield  {author} {\bibinfo {author} {\bibfnamefont {A.}~\bibnamefont
  {Gallem{\'\i}}}, \bibinfo {author} {\bibfnamefont {L.~P.}\ \bibnamefont
  {Pitaevskii}}, \bibinfo {author} {\bibfnamefont {S.}~\bibnamefont
  {Stringari}}, \ and\ \bibinfo {author} {\bibfnamefont {A.}~\bibnamefont
  {Recati}},\ }\href@noop {} {\bibfield  {journal} {\bibinfo  {journal} {Phys.
  Rev. A}\ }\textbf {\bibinfo {volume} {97}},\ \bibinfo {pages} {063615}
  (\bibinfo {year} {2018})}\BibitemShut {NoStop}%
\bibitem [{\citenamefont {Kobayashi}\ \emph {et~al.}(2018)\citenamefont
  {Kobayashi}, \citenamefont {Eto},\ and\ \citenamefont
  {Nitta}}]{kobayashi2018berezinskii}%
  \BibitemOpen
  \bibfield  {author} {\bibinfo {author} {\bibfnamefont {M.}~\bibnamefont
  {Kobayashi}}, \bibinfo {author} {\bibfnamefont {M.}~\bibnamefont {Eto}}, \
  and\ \bibinfo {author} {\bibfnamefont {M.}~\bibnamefont {Nitta}},\
  }\href@noop {} {\bibfield  {journal} {\bibinfo  {journal} {arXiv:1802.08763}\
  } (\bibinfo {year} {2018})}\BibitemShut {NoStop}%
\bibitem [{\citenamefont {Babaev}(2004)}]{babaev2004andreev}%
  \BibitemOpen
  \bibfield  {author} {\bibinfo {author} {\bibfnamefont {E.}~\bibnamefont
  {Babaev}},\ }\href {\doibase 10.1103/PhysRevD.70.043001} {\bibfield
  {journal} {\bibinfo  {journal} {Phys. Rev. D}\ }\textbf {\bibinfo {volume}
  {70}},\ \bibinfo {pages} {043001} (\bibinfo {year} {2004})}\BibitemShut
  {NoStop}%
\bibitem [{\citenamefont {Babaev}\ \emph {et~al.}(2005)\citenamefont {Babaev},
  \citenamefont {Sudb{\o}},\ and\ \citenamefont
  {Ashcroft}}]{babaev2005observability}%
  \BibitemOpen
  \bibfield  {author} {\bibinfo {author} {\bibfnamefont {E.}~\bibnamefont
  {Babaev}}, \bibinfo {author} {\bibfnamefont {A.}~\bibnamefont {Sudb{\o}}}, \
  and\ \bibinfo {author} {\bibfnamefont {N.~W.}\ \bibnamefont {Ashcroft}},\
  }\href {\doibase 10.1103/PhysRevLett.95.105301} {\bibfield  {journal}
  {\bibinfo  {journal} {Phys. Rev. Lett.}\ }\textbf {\bibinfo {volume} {95}},\
  \bibinfo {pages} {105301} (\bibinfo {year} {2005})}\BibitemShut {NoStop}%
\bibitem [{\citenamefont {{LeClair}}\ \emph {et~al.}(1998)\citenamefont
  {{LeClair}}, \citenamefont {{Ludwig}},\ and\ \citenamefont
  {{Mussardo}}}]{LeClair1998}%
  \BibitemOpen
  \bibfield  {author} {\bibinfo {author} {\bibfnamefont {A.}~\bibnamefont
  {{LeClair}}}, \bibinfo {author} {\bibfnamefont {A.~W.~W.}\ \bibnamefont
  {{Ludwig}}}, \ and\ \bibinfo {author} {\bibfnamefont {G.}~\bibnamefont
  {{Mussardo}}},\ }\href {\doibase 10.1016/S0550-3213(97)00724-4} {\bibfield
  {journal} {\bibinfo  {journal} {Nucl. Phys. B}\ }\textbf {\bibinfo {volume}
  {512}},\ \bibinfo {pages} {523} (\bibinfo {year} {1998})}\BibitemShut
  {NoStop}%
\bibitem [{\citenamefont {Stewart}(1984)}]{Stewart1984}%
  \BibitemOpen
  \bibfield  {author} {\bibinfo {author} {\bibfnamefont {G.~R.}\ \bibnamefont
  {Stewart}},\ }\href {\doibase 10.1103/RevModPhys.56.755} {\bibfield
  {journal} {\bibinfo  {journal} {Rev. Mod. Phys.}\ }\textbf {\bibinfo {volume}
  {56}},\ \bibinfo {pages} {755} (\bibinfo {year} {1984})}\BibitemShut
  {NoStop}%
\bibitem [{\citenamefont {Pavarini}\ \emph {et~al.}(2015)\citenamefont
  {Pavarini}, \citenamefont {Koch},\ and\ \citenamefont
  {Coleman}}]{Coleman2015}%
  \BibitemOpen
  \bibfield  {author} {\bibinfo {author} {\bibfnamefont {E.}~\bibnamefont
  {Pavarini}}, \bibinfo {author} {\bibfnamefont {E.}~\bibnamefont {Koch}}, \
  and\ \bibinfo {author} {\bibfnamefont {P.}~\bibnamefont {Coleman}},\
  }\href@noop {} {\emph {\bibinfo {title} {{M}any-{B}ody {P}hysics: {F}rom
  {K}ondo to {H}ubbard}}}\ (\bibinfo  {publisher} {Forschungszentrum Jülich
  GmbH Zentralbibliothek, Verlag},\ \bibinfo {year} {2015})\ \bibinfo {note}
  {{C}hap.1: Heavy Fermions and the Kondo Lattice: A 21st Century
  Perspective}\BibitemShut {NoStop}%
\bibitem [{\citenamefont {{Strong}}\ and\ \citenamefont
  {{Millis}}(1994)}]{Strong1994}%
  \BibitemOpen
  \bibfield  {author} {\bibinfo {author} {\bibfnamefont {S.~P.}\ \bibnamefont
  {{Strong}}}\ and\ \bibinfo {author} {\bibfnamefont {A.~J.}\ \bibnamefont
  {{Millis}}},\ }\href {\doibase 10.1103/PhysRevB.50.9911} {\bibfield
  {journal} {\bibinfo  {journal} {Phys. Rev. B}\ }\textbf {\bibinfo {volume}
  {50}},\ \bibinfo {pages} {9911} (\bibinfo {year} {1994})}\BibitemShut
  {NoStop}%
\bibitem [{\citenamefont {Po}\ \emph {et~al.}(2018)\citenamefont {Po},
  \citenamefont {Zou}, \citenamefont {Vishwanath},\ and\ \citenamefont
  {Senthil}}]{Po2018}%
  \BibitemOpen
  \bibfield  {author} {\bibinfo {author} {\bibfnamefont {H.~C.}\ \bibnamefont
  {Po}}, \bibinfo {author} {\bibfnamefont {L.}~\bibnamefont {Zou}}, \bibinfo
  {author} {\bibfnamefont {A.}~\bibnamefont {Vishwanath}}, \ and\ \bibinfo
  {author} {\bibfnamefont {T.}~\bibnamefont {Senthil}},\ }\href {\doibase
  10.1103/PhysRevX.8.031089} {\bibfield  {journal} {\bibinfo  {journal} {Phys.
  Rev. X}\ }\textbf {\bibinfo {volume} {8}},\ \bibinfo {pages} {031089}
  (\bibinfo {year} {2018})}\BibitemShut {NoStop}%
\bibitem [{\citenamefont {Ramires}\ and\ \citenamefont
  {Lado}(2018)}]{Ramires2018}%
  \BibitemOpen
  \bibfield  {author} {\bibinfo {author} {\bibfnamefont {A.}~\bibnamefont
  {Ramires}}\ and\ \bibinfo {author} {\bibfnamefont {J.~L.}\ \bibnamefont
  {Lado}},\ }\href {\doibase 10.1103/PhysRevLett.121.146801} {\bibfield
  {journal} {\bibinfo  {journal} {Phys. Rev. Lett.}\ }\textbf {\bibinfo
  {volume} {121}},\ \bibinfo {pages} {146801} (\bibinfo {year}
  {2018})}\BibitemShut {NoStop}%
\bibitem [{\citenamefont {Peltonen}\ \emph {et~al.}(2018)\citenamefont
  {Peltonen}, \citenamefont {Ojaj\"arvi},\ and\ \citenamefont
  {Heikkil\"a}}]{Peltonen2018}%
  \BibitemOpen
  \bibfield  {author} {\bibinfo {author} {\bibfnamefont {T.~J.}\ \bibnamefont
  {Peltonen}}, \bibinfo {author} {\bibfnamefont {R.}~\bibnamefont
  {Ojaj\"arvi}}, \ and\ \bibinfo {author} {\bibfnamefont {T.~T.}\ \bibnamefont
  {Heikkil\"a}},\ }\href {\doibase 10.1103/PhysRevB.98.220504} {\bibfield
  {journal} {\bibinfo  {journal} {Phys. Rev. B}\ }\textbf {\bibinfo {volume}
  {98}},\ \bibinfo {pages} {220504} (\bibinfo {year} {2018})}\BibitemShut
  {NoStop}%
\bibitem [{\citenamefont {{Cao}}\ \emph {et~al.}(2018)\citenamefont {{Cao}},
  \citenamefont {{Fatemi}}, \citenamefont {{Fang}}, \citenamefont {{Watanabe}},
  \citenamefont {{Taniguchi}}, \citenamefont {{Kaxiras}},\ and\ \citenamefont
  {{Jarillo-Herrero}}}]{Cao2018}%
  \BibitemOpen
  \bibfield  {author} {\bibinfo {author} {\bibfnamefont {Y.}~\bibnamefont
  {{Cao}}}, \bibinfo {author} {\bibfnamefont {V.}~\bibnamefont {{Fatemi}}},
  \bibinfo {author} {\bibfnamefont {S.}~\bibnamefont {{Fang}}}, \bibinfo
  {author} {\bibfnamefont {K.}~\bibnamefont {{Watanabe}}}, \bibinfo {author}
  {\bibfnamefont {T.}~\bibnamefont {{Taniguchi}}}, \bibinfo {author}
  {\bibfnamefont {E.}~\bibnamefont {{Kaxiras}}}, \ and\ \bibinfo {author}
  {\bibfnamefont {P.}~\bibnamefont {{Jarillo-Herrero}}},\ }\href {\doibase
  10.1038/nature26160} {\bibfield  {journal} {\bibinfo  {journal} {\nat}\
  }\textbf {\bibinfo {volume} {556}},\ \bibinfo {pages} {43} (\bibinfo {year}
  {2018})}\BibitemShut {NoStop}%
\bibitem [{\citenamefont {{Leggett}}(2006)}]{Leggett2006}%
  \BibitemOpen
  \bibfield  {author} {\bibinfo {author} {\bibfnamefont {A.~J.}\ \bibnamefont
  {{Leggett}}},\ }\href {\doibase 10.1038/nphys254} {\bibfield  {journal}
  {\bibinfo  {journal} {Nature Phys.}\ }\textbf {\bibinfo {volume} {2}},\
  \bibinfo {pages} {134} (\bibinfo {year} {2006})}\BibitemShut {NoStop}%
\bibitem [{\citenamefont {{Pierson}}(1994)}]{Pierson1994}%
  \BibitemOpen
  \bibfield  {author} {\bibinfo {author} {\bibfnamefont {S.~W.}\ \bibnamefont
  {{Pierson}}},\ }\href {\doibase 10.1103/PhysRevLett.73.2496} {\bibfield
  {journal} {\bibinfo  {journal} {\prl}\ }\textbf {\bibinfo {volume} {73}},\
  \bibinfo {pages} {2496} (\bibinfo {year} {1994})}\BibitemShut {NoStop}%
\bibitem [{\citenamefont {{N{\'a}ndori}}\ \emph {et~al.}(2005)\citenamefont
  {{N{\'a}ndori}}, \citenamefont {{Nagy}}, \citenamefont {{Sailer}},\ and\
  \citenamefont {{Jentschura}}}]{Nandori2005}%
  \BibitemOpen
  \bibfield  {author} {\bibinfo {author} {\bibfnamefont {I.}~\bibnamefont
  {{N{\'a}ndori}}}, \bibinfo {author} {\bibfnamefont {S.}~\bibnamefont
  {{Nagy}}}, \bibinfo {author} {\bibfnamefont {K.}~\bibnamefont {{Sailer}}}, \
  and\ \bibinfo {author} {\bibfnamefont {U.~D.}\ \bibnamefont {{Jentschura}}},\
  }\href {\doibase 10.1016/j.nuclphysb.2005.07.016} {\bibfield  {journal}
  {\bibinfo  {journal} {Nucl. Phys. B}\ }\textbf {\bibinfo {volume} {725}},\
  \bibinfo {pages} {467} (\bibinfo {year} {2005})}\BibitemShut {NoStop}%
\bibitem [{\citenamefont {Mathey}\ \emph {et~al.}(2008)\citenamefont {Mathey},
  \citenamefont {Polkovnikov},\ and\ \citenamefont {{Castro
  Neto}}}]{mathey2008}%
  \BibitemOpen
  \bibfield  {author} {\bibinfo {author} {\bibfnamefont {L.}~\bibnamefont
  {Mathey}}, \bibinfo {author} {\bibfnamefont {A.}~\bibnamefont {Polkovnikov}},
  \ and\ \bibinfo {author} {\bibfnamefont {A.~H.}\ \bibnamefont {{Castro
  Neto}}},\ }\href {\doibase 10.1209/0295-5075/81/10008} {\bibfield  {journal}
  {\bibinfo  {journal} {Europhys. Lett.}\ }\textbf {\bibinfo {volume} {81}},\
  \bibinfo {pages} {10008} (\bibinfo {year} {2008})}\BibitemShut {NoStop}%
\bibitem [{\citenamefont {{Pierson}}(1995)}]{Pierson1995}%
  \BibitemOpen
  \bibfield  {author} {\bibinfo {author} {\bibfnamefont {S.~W.}\ \bibnamefont
  {{Pierson}}},\ }\href {\doibase 10.1103/PhysRevLett.74.2359} {\bibfield
  {journal} {\bibinfo  {journal} {\prl}\ }\textbf {\bibinfo {volume} {74}},\
  \bibinfo {pages} {2359} (\bibinfo {year} {1995})}\BibitemShut {NoStop}%
\bibitem [{\citenamefont {{N{\'a}ndori}}\ \emph {et~al.}(2007)\citenamefont
  {{N{\'a}ndori}}, \citenamefont {{Jentschura}}, \citenamefont {{Nagy}},
  \citenamefont {{Sailer}}, \citenamefont {{Vad}},\ and\ \citenamefont
  {{M{\'e}sz{\'a}ros}}}]{Nandori2007}%
  \BibitemOpen
  \bibfield  {author} {\bibinfo {author} {\bibfnamefont {I.}~\bibnamefont
  {{N{\'a}ndori}}}, \bibinfo {author} {\bibfnamefont {U.~D.}\ \bibnamefont
  {{Jentschura}}}, \bibinfo {author} {\bibfnamefont {S.}~\bibnamefont
  {{Nagy}}}, \bibinfo {author} {\bibfnamefont {K.}~\bibnamefont {{Sailer}}},
  \bibinfo {author} {\bibfnamefont {K.}~\bibnamefont {{Vad}}}, \ and\ \bibinfo
  {author} {\bibfnamefont {S.}~\bibnamefont {{M{\'e}sz{\'a}ros}}},\ }\href
  {\doibase 10.1088/0953-8984/19/23/236226} {\bibfield  {journal} {\bibinfo
  {journal} {J. Phys. Cond. Mat.}\ }\textbf {\bibinfo {volume} {19}},\ \bibinfo
  {eid} {236226} (\bibinfo {year} {2007})}\BibitemShut {NoStop}%
\bibitem [{\citenamefont {Adhikari}(1986)}]{adhikari1986quantum}%
  \BibitemOpen
  \bibfield  {author} {\bibinfo {author} {\bibfnamefont {S.~K.}\ \bibnamefont
  {Adhikari}},\ }\href {\doibase 10.1119/1.14623} {\bibfield  {journal}
  {\bibinfo  {journal} {Am. J. Phys.}\ }\textbf {\bibinfo {volume} {54}},\
  \bibinfo {pages} {362} (\bibinfo {year} {1986})}\BibitemShut {NoStop}%
\bibitem [{\citenamefont {Petrov}\ \emph
  {et~al.}(2000{\natexlab{b}})\citenamefont {Petrov}, \citenamefont
  {Holzmann},\ and\ \citenamefont {Shlyapnikov}}]{petrov2000bose}%
  \BibitemOpen
  \bibfield  {author} {\bibinfo {author} {\bibfnamefont {D.~S.}\ \bibnamefont
  {Petrov}}, \bibinfo {author} {\bibfnamefont {M.}~\bibnamefont {Holzmann}}, \
  and\ \bibinfo {author} {\bibfnamefont {G.~V.}\ \bibnamefont {Shlyapnikov}},\
  }\href@noop {} {\bibfield  {journal} {\bibinfo  {journal} {Phys. Rev. Lett.}\
  }\textbf {\bibinfo {volume} {84}},\ \bibinfo {pages} {2551} (\bibinfo {year}
  {2000}{\natexlab{b}})}\BibitemShut {NoStop}%
\bibitem [{\citenamefont {Petrov}\ and\ \citenamefont
  {Shlyapnikov}(2001)}]{petrov2001interatomic}%
  \BibitemOpen
  \bibfield  {author} {\bibinfo {author} {\bibfnamefont {D.~S.}\ \bibnamefont
  {Petrov}}\ and\ \bibinfo {author} {\bibfnamefont {G.~V.}\ \bibnamefont
  {Shlyapnikov}},\ }\href {\doibase 10.1103/PhysRevA.64.012706} {\bibfield
  {journal} {\bibinfo  {journal} {Phys. Rev. A}\ }\textbf {\bibinfo {volume}
  {64}},\ \bibinfo {pages} {012706} (\bibinfo {year} {2001})}\BibitemShut
  {NoStop}%
\bibitem [{\citenamefont {Salasnich}\ and\ \citenamefont
  {Toigo}(2016)}]{salasnich2016zero}%
  \BibitemOpen
  \bibfield  {author} {\bibinfo {author} {\bibfnamefont {L.}~\bibnamefont
  {Salasnich}}\ and\ \bibinfo {author} {\bibfnamefont {F.}~\bibnamefont
  {Toigo}},\ }\href@noop {} {\bibfield  {journal} {\bibinfo  {journal} {Phys.
  Rep.}\ }\textbf {\bibinfo {volume} {640}},\ \bibinfo {pages} {1} (\bibinfo
  {year} {2016})}\BibitemShut {NoStop}%
\bibitem [{\citenamefont {Prokof'ev}\ \emph {et~al.}(2004)\citenamefont
  {Prokof'ev}, \citenamefont {Ruebenacker},\ and\ \citenamefont
  {Svistunov}}]{Prokofiev2004}%
  \BibitemOpen
  \bibfield  {author} {\bibinfo {author} {\bibfnamefont {N.}~\bibnamefont
  {Prokof'ev}}, \bibinfo {author} {\bibfnamefont {O.}~\bibnamefont
  {Ruebenacker}}, \ and\ \bibinfo {author} {\bibfnamefont {B.}~\bibnamefont
  {Svistunov}},\ }\href {\doibase 10.1103/PhysRevA.69.053625} {\bibfield
  {journal} {\bibinfo  {journal} {Phys. Rev. A}\ }\textbf {\bibinfo {volume}
  {69}},\ \bibinfo {pages} {053625} (\bibinfo {year} {2004})}\BibitemShut
  {NoStop}%
\bibitem [{\citenamefont {Landau}\ and\ \citenamefont
  {Lifshitz}(1980)}]{landau1980course}%
  \BibitemOpen
  \bibfield  {author} {\bibinfo {author} {\bibfnamefont {L.~D.}\ \bibnamefont
  {Landau}}\ and\ \bibinfo {author} {\bibfnamefont {E.~M.}\ \bibnamefont
  {Lifshitz}},\ }\href@noop {} {\emph {\bibinfo {title} {Course of Theoretical
  Physics. Vol. 9: Statistical Physics}}}\ (\bibinfo  {publisher} {Pergamon
  Press, Oxford},\ \bibinfo {year} {1980})\BibitemShut {NoStop}%
\bibitem [{\citenamefont {Abad}\ \emph {et~al.}(2014)\citenamefont {Abad},
  \citenamefont {Sartori}, \citenamefont {Finazzi},\ and\ \citenamefont
  {Recati}}]{abad2014persistent}%
  \BibitemOpen
  \bibfield  {author} {\bibinfo {author} {\bibfnamefont {M.}~\bibnamefont
  {Abad}}, \bibinfo {author} {\bibfnamefont {A.}~\bibnamefont {Sartori}},
  \bibinfo {author} {\bibfnamefont {S.}~\bibnamefont {Finazzi}}, \ and\
  \bibinfo {author} {\bibfnamefont {A.}~\bibnamefont {Recati}},\ }\href@noop {}
  {\bibfield  {journal} {\bibinfo  {journal} {Phys. Rev. A}\ }\textbf {\bibinfo
  {volume} {89}},\ \bibinfo {pages} {053602} (\bibinfo {year}
  {2014})}\BibitemShut {NoStop}%
\bibitem [{\citenamefont {Defenu}\ \emph {et~al.}(2017)\citenamefont {Defenu},
  \citenamefont {Trombettoni}, \citenamefont {N{\'a}ndori},\ and\ \citenamefont
  {Enss}}]{defenu2017nonperturbative}%
  \BibitemOpen
  \bibfield  {author} {\bibinfo {author} {\bibfnamefont {N.}~\bibnamefont
  {Defenu}}, \bibinfo {author} {\bibfnamefont {A.}~\bibnamefont {Trombettoni}},
  \bibinfo {author} {\bibfnamefont {I.}~\bibnamefont {N{\'a}ndori}}, \ and\
  \bibinfo {author} {\bibfnamefont {T.}~\bibnamefont {Enss}},\ }\href {\doibase
  10.1103/PhysRevB.96.174505} {\bibfield  {journal} {\bibinfo  {journal} {Phys.
  Rev. B}\ }\textbf {\bibinfo {volume} {96}},\ \bibinfo {pages} {174505}
  (\bibinfo {year} {2017})}\BibitemShut {NoStop}%
\bibitem [{\citenamefont {{Villain}}(1975)}]{Villain1975}%
  \BibitemOpen
  \bibfield  {author} {\bibinfo {author} {\bibfnamefont {J.}~\bibnamefont
  {{Villain}}},\ }\href {\doibase 10.1016/0378-4363(75)90101-1} {\bibfield
  {journal} {\bibinfo  {journal} {Physica B}\ }\textbf {\bibinfo {volume}
  {79}},\ \bibinfo {pages} {1} (\bibinfo {year} {1975})}\BibitemShut {NoStop}%
\bibitem [{\citenamefont {Jos{\'e}}\ \emph {et~al.}(1977)\citenamefont
  {Jos{\'e}}, \citenamefont {Kadanoff}, \citenamefont {Kirkpatrick},\ and\
  \citenamefont {Nelson}}]{jose1977renormalization}%
  \BibitemOpen
  \bibfield  {author} {\bibinfo {author} {\bibfnamefont {J.~V.}\ \bibnamefont
  {Jos{\'e}}}, \bibinfo {author} {\bibfnamefont {L.~P.}\ \bibnamefont
  {Kadanoff}}, \bibinfo {author} {\bibfnamefont {S.}~\bibnamefont
  {Kirkpatrick}}, \ and\ \bibinfo {author} {\bibfnamefont {D.~R.}\ \bibnamefont
  {Nelson}},\ }\href@noop {} {\bibfield  {journal} {\bibinfo  {journal} {Phys.
  Rev. B}\ }\textbf {\bibinfo {volume} {16}},\ \bibinfo {pages} {1217}
  (\bibinfo {year} {1977})}\BibitemShut {NoStop}%
\bibitem [{Note1()}]{Note1}%
  \BibitemOpen
  \bibinfo {note} {Vortices with winding number $2n$ have larger energy than
  two vortices with winding number $n$. In the low-temperature limit, vortices
  of winding numbers $|w_j|>1$ are unstable with respect to the decay into
  vortices of smaller winding numbers~\cite {svistunov2015superfluid}.
  Therefore, the low-temperature phase is dominated by $w_j=\pm 1$
  excitations.}\BibitemShut {Stop}%
\bibitem [{\citenamefont {Altland}\ and\ \citenamefont
  {Simons}(2010)}]{altland2010condensed}%
  \BibitemOpen
  \bibfield  {author} {\bibinfo {author} {\bibfnamefont {A.}~\bibnamefont
  {Altland}}\ and\ \bibinfo {author} {\bibfnamefont {B.~D.}\ \bibnamefont
  {Simons}},\ }\href@noop {} {\emph {\bibinfo {title} {Condensed Matter Field
  Theory}}}\ (\bibinfo  {publisher} {Cambridge University Press},\ \bibinfo
  {year} {2010})\BibitemShut {NoStop}%
\bibitem [{Note2()}]{Note2}%
  \BibitemOpen
  \bibinfo {note} {However, since \protect \textup {\hbox {\mathsurround \z@
  \protect \normalfont (\ignorespaces \ref {eq:Ls}\unskip \@@italiccorr )}} can
  be extended straightforwardly to the case of different masses, the RG flow is
  valid also in that case.}\BibitemShut {Stop}%
\bibitem [{Note3()}]{Note3}%
  \BibitemOpen
  \bibinfo {note} {Mixed vortex contributions turn up only at fourth order due
  to charge neutrality, see Appendix D.}\BibitemShut {Stop}%
\bibitem [{\citenamefont {Weinberg}(1995)}]{weinberg1995quantum}%
  \BibitemOpen
  \bibfield  {author} {\bibinfo {author} {\bibfnamefont {S.}~\bibnamefont
  {Weinberg}},\ }\href@noop {} {\emph {\bibinfo {title} {The Quantum Theory of
  Fields}}},\ Vol.~\bibinfo {volume} {2}\ (\bibinfo  {publisher} {Cambridge
  University Press},\ \bibinfo {year} {1995})\BibitemShut {NoStop}%
\end{thebibliography}%
\end{document}